\documentclass{aa}  

\usepackage{gensymb}

\usepackage{amsmath}

\usepackage[T1]{fontenc}

\DeclareRobustCommand{\VAN}[3]{#2}
\let\VANthebibliography\thebibliography
\def\thebibliography{\DeclareRobustCommand{\VAN}[3]{##3}\VANthebibliography}

\usepackage{amssymb}	
\usepackage{graphicx}	
\usepackage{txfonts}
\usepackage{titlesec}
\usepackage{multibib}
\usepackage[dvipsnames]{xcolor}
\usepackage{enumitem}
\usepackage{fancyhdr}
\usepackage[nointegrals]{wasysym} 
\usepackage{subcaption}
\usepackage{orcidlink}
\usepackage[graphicx]{realboxes} 
\usepackage{pifont}

\usepackage{placeins}  

{

\begin{document}
\title{The \emph{clus} model in SPEX: projection and resonant scattering effects on the iron abundance and temperature profiles of galaxy clusters}
\titlerunning{SPEX: the clus model}
\author{L{\' y}dia {\v S}tofanov{\' a} \inst{1,2,3}\orcidlink{0000-0003-0049-6205}, Aurora Simionescu \inst{2,1,4}\orcidlink{0000-0002-9714-3862}, and  Jelle S. Kaastra \inst{2,1}\orcidlink{0000-0001-5540-2822}    }
\institute{Leiden Observatory, Leiden University, PO Box 9513, 2300RA, Leiden, The Netherlands\\
			\email{stofanova@physics.muni.cz}
			\and SRON Netherlands Institute for Space Research, Niels Bohrweg 4, 2333 CA Leiden, The Netherlands
			\and Department of Theoretical Physics and Astrophysics, Faculty of Science, Masaryk University, Kotlá{\v r}sk{\' a} 2, Brno 611 37, Czech Republic
			\and Kavli Institute for the Physics and Mathematics of the Universe, The University of Tokyo, Kashiwa, Chiba 277-8583, Japan}
\authorrunning{{\v S}tofanov{\' a} et al.}
\date{Accepted XXX. Received YYY; in original form ZZZ}

\abstract{
In this paper we introduce the \emph{clus} model, which has been newly implemented in the X-ray spectral fitting software package SPEX. Based on the 3D radial profiles of the gas density, temperature, metal abundance, turbulent, and inflow/outflow velocities, the \emph{clus} model creates spectra for a chosen projected region on the sky. Additionally, it can also take into account the resonant scattering. We show a few applications of the \emph{clus} model on simulated spectra of the massive elliptical galaxy NGC 4636, and galaxy clusters A383, A2029, A1795, A262, and the Perseus cluster. We quantify the effect of projection, as well as resonant scattering on inferred profiles of the iron abundance and temperature, assuming the resolution similar to Chandra ACIS-S and XRISM Resolve. Our results show that, depending on the mass of the object, as well as the projected distance from its core, neither a single-temperature, double-temperature, nor the Gaussian-shaped differential emission measure models can accurately describe the input emission measure distribution of these massive objects. The largest effect of projection as well as resonant scattering is seen for projected profiles of iron abundance of NGC 4636, where we are able to reproduce the observed iron abundance drop in its inner-most few kiloparsecs. Furthermore, we find that projection effects also influence the best-fit temperature, and the magnitude of this effect varies depending on the underlying hydrodynamical profiles of individual objects. In the core, the projection effects are the largest for A1795 and NGC 4636, while in the outskirts the largest difference between 2D and 3D temperature profiles are for Perseus and A1795, regardless of the instrumental resolution. These findings might potentially have an impact on cross-calibration studies between different instruments, as well as on the precision cosmology. 

}


\maketitle

\section{Introduction}
\label{Sec:intro}

Galaxy clusters, galaxy groups, and massive elliptical galaxies appear as diffused sources of light at X-ray wavelengths \citep{1972ApJ...178..309F, 1972ApJ...174L..65K}. Plasma revealed by this X-ray emission is referred to as the intra-cluster medium (ICM), the intra-group medium (IGrM), or the circum-galactic medium (CGM), and forms a major fraction of the baryonic matter in these objects. It emits mostly through the free-free continuum (thermal bremsstrahlung), and through line emission. The X-ray spectroscopy of these sources can be used to derive many of their properties, as for example their electron temperature, electron density, turbulent velocities, or metal abundances \citep{2010A&ARv..18..127B}. 

In the last few decades, the capabilities of X-ray spectroscopy were showcased mainly through CCD and grating spectrometers. The Chandra X-ray Observatory, XMM-Newton, and Suzaku allowed great progress thanks to their improved spectral and spatial resolution, as well as larger effective area and low background. Another milestone that X-ray spectroscopy reached was in 2016, when a micro-calorimeter was launched on board of the Hitomi satellite, which offered an unprecedented spectral resolution of $5$\,eV at the energy of $6$\,keV.  XRISM, which was launched to its orbit in September 2023, and which carries on the legacy of Hitomi, will continue observing the Universe through high-resolution spectroscopy.  

Among many discoveries enabled with Chandra, XMM-Newton, and Suzaku, these observatories provided detailed measurements of the metal distribution in galaxy clusters, galaxy groups, and giant elliptical galaxies. Many of these objects show flattened, or peaky metal (mostly iron) abundance profiles towards their cores, however, X-ray observations of some of them revealed a steep decrease (sometimes by as much as $50$\% of their global maximum) of metal abundances in the few innermost kiloparsecs \citep{2003ApJ...590..225C, 2002MNRAS.336..299J,  2002MNRAS.331..273S, 2007MNRAS.381.1381S, 2010MNRAS.405.1624M, 2013MNRAS.428...58R, 2017A&A...603A..80M}.

A few possible scenarios have arisen to explain these central abundance drops, but their origin has not yet been established. For example, metals can be transported to larger distances by the mechanical energy powered by the active galactic nucleus (AGN) feedback \citep{2008A&A...482...97S, 2009A&A...493..409S}, or by galactic winds \citep{2005A&A...435L..25S}. Studies by  \citet{2013MNRAS.433.3290P, 2015MNRAS.447..417P} suggested instead that iron might get depleted into dust grains in the cores of these massive objects, and therefore it becomes 'invisible' in X-rays. If this scenario is true, and if the elements such as iron, silicon, sulphur, magnesium or calcium would show the central abundance drop, the noble gasses such as neon or argon should be indifferent to this process. Such behaviour has been confirmed with Chandra in \citet{2019A&A...623A..17L}, who reported that in the central 10\,kpc of the Centaurus galaxy cluster, elements such as Fe, Si, S, Mg, and Ca show an abundance drop, however, no abundance drop was found in Ar. \citet{2019MNRAS.485.1651L} concluded similar results for $10$ cool-core clusters. Contrary to these conclusions, measurements with XMM-Newton suggested that the observed abundance drop in the Centaurus cluster could be partly caused by systematic uncertainties in the atomic data and response matrices, rather than the depletion of iron into dust \citep{2022MNRAS.514.4222F}. Future XRISM observations of metals in the Centaurus galaxy cluster promise to resolve this mystery. Other scenarios originate from incompleteness in the spectral fitting models, as for example projection effects \citep{2007MNRAS.381.1381S}, modelling of the central brightest cluster galaxy (BCG), or a simplification of the multi-temperature plasma with a single-temperature fitting model, which is also known as the ``iron bias'' \citep{2000MNRAS.311..176B}. Another scenario suggests sedimentation of helium nuclei in the cluster cores, which could potentially raise the helium abundance values above the solar one, and therefore overestimate the continuum, and underestimate all other abundances \citep{2006MNRAS.369L..42E}. This scenario is, however, impossible to test directly in X-rays, since He does not have transitions in X-ray wavelengths. 

An alternative explanation arose from an idea proposed in \citet{1987SvAL...13....3G}, which shows that the radial X-ray surface brightness profile of galaxy clusters can be distorted by the scattering of X-ray photons in the resonant lines of highly charged and heavy-metal ions. Some lines in the ICM have optical depths larger than unity, and undergo the resonant scattering (RS) process. If a photon is resonantly scattered, it gets absorbed and almost immediately re-emitted in a different direction. This causes the flux in the resonant line to drop in the cluster cores, however, in the outskirts the flux of the same resonant line is enhanced. Since the optical depth of each spectral line depends, among other things, also on the turbulent velocity, RS has been used as an independent proxy to measure gas motions and turbulent velocities in the ICM and IGrM, in addition to the gas velocities obtained from the line broadening (see e.g. \citealp{2004MNRAS.347...29C, 2009MNRAS.398...23W, 2012A&A...539A..34D, 2013MNRAS.435.3111Z, 2017MNRAS.472.1659O, 2018SSRv..214..108G}). Recent study of \citet{2023MNRAS.522.3665N} showed that RS could be used to observe CGM around galaxies with future X-ray observatories. Additionally, RS has been studied in relation to the abundance drop in the Centaurus galaxy cluster, however, \citet{2006MNRAS.370...63S} showed that accounting for RS does not remove the central dip in its abundance profile. In case of the massive elliptical galaxy NGC 4636, \citet{2009MNRAS.398...23W} concluded that neglecting the effect of RS underestimates the chemical abundances of iron and oxygen by $\sim 10 - 20$\%. 

In this paper we present the \emph{clus} model, which has been newly implemented in the SPEctral X-ray and EUV (SPEX) software package, and which calculates the spectrum and radial profiles for a spherically symmetric approximation of a cluster or a group of galaxies, or even massive elliptical galaxies. In section\,\ref{Sec:methods} we describe details of this model, our sample selection, and the fitting procedure. In section\,\ref{Sec:cluster_model_application} we provide an application of the \emph{clus} model to the Chandra ACIS-S and XRISM Resolve simulated data, and we address whether the \emph{clus} model can help with understanding the origin of the iron abundance drops in the cores of some galaxy clusters, groups, or giant elliptical galaxies. We also discuss the effects on the projected temperature profiles. In section\,\ref{Sec:discussion} we discuss our results and compare our findings to existing literature studies. Lastly, section\,\ref{Sec:conclusions} summarizes our main conclusions.  

\section{Methods}
\label{Sec:methods}

\subsection{SPEX}
We use the SPEX software package \citep{kaastra1996_spex, kaastra2018_spex, kaastra_j_s_2020_4384188} in version $3.08$\footnote{For the most recent version see \url{https://spex-xray.github.io/spex-help/changelog.html}.} for modelling and analysis of high-resolution X-ray spectra. With its own atomic database SPEXACT (The SPEX Atomic Code \& Tables) it includes around $4.2 \times 10^6$ lines from $30$ different chemical elements (H to Zn). Since the first version of SPEX, many updates have been made, including the implementation of the  \emph{clus} model which is described in more details in this paper. Unless stated otherwise, we use the proto-solar abundances by \citet{2009LanB...4B..712L}. 

\subsection{The \emph{clus} model}
\label{Sec:cluster_model_theory}

The \emph{clus} model takes as input parametrised 3D radial profiles of the gas density, temperature, metal abundance, turbulent, as well as inflow/outflow velocities. Given these parametrised profiles, the emission in multiple 3D shells, each approximated as a single temperature model in collisional ionisation equilibrium (CIE), is computed and projected onto the sky. The number of shells can be adjusted and is set by a user through the parameters $nr$ (number of 3D shells for which the model is evaluated) and $npro$ (number of projected annuli). The 3D model is evaluated on a logarithmically spaced grid, which is defined by the inner radius $r_{\mathrm{in}}$ and outer radius $r_{\mathrm{out}}$, where at radius $r_{\mathrm{out}}$ the density is set to zero. Radius $r_{\mathrm{in}}$ is taken as $1\%$ of $r_{500}$, or if it is smaller, $10\%$ of the smallest of the core radii in the density profile (see Sec.\,\ref{Sec:density_profile}). For the innermost shell, the inner boundary $r_{\mathrm{in}}$ is replaced by zero. The \emph{clus} model can then be run in different modes, where the output is either the spectral energy distribution within a user-defined projected spatial region, or a radial surface brightness profile in a user-requested energy band. 

More specifically, the assumed shapes of the underlying 3D profiles used as input to the \emph{clus} model follow:
\begin{itemize}[leftmargin=11pt]
	\item a two $\beta$ model for the density distribution (with the additional possibility to introduce a jump or a break in the slope at a given radius, see Sec.\,\ref{Sec:density_profile}),
	\item the same functional form for the temperature profile as that proposed in \citet{2006ApJ...640..691V}, again with a possible additional jump (see Sec.\,\ref{Sec:temperature_profile}),
	\item the universal abundance profile as defined in \citet{2017A&A...603A..80M} (see Sec.\,\ref{Sec:abundance_profile}).
\end{itemize}

The hydrodynamical profiles such as density or temperature are evaluated for the middle radius of each shell using analytical parametrization of the 3D profiles, and it is assumed that these quantities are constant within the shell. Therefore, the emission measure of each shell is simply evaluated as the product of electron and hydrogen density at the central radius of the shell, multiplied by the volume of the shell. In case the user introduces a density discontinuity in the model, the radius of the density discontinuity is also chosen as an additional grid point (if it is close to an existing grid point, that existing grid point is moved to the discontinuity).

The \emph{clus} model takes a linear grid for the projected 2D profiles, which emulates real detectors that are designed with a linear scale. For the 3D profiles, the quantities are evaluated at the shell central radii, while for the 2D projected profiles the quantities are evaluated at sub-shell resolution (typically using $20$ points for each shell). Within the shell it is assumed that the emission measure as a function of radius follows a power-law. The power-law slope is determined by comparing the emission measures of the neighbouring shells.

The \emph{clus} model also has two ascii-output options: (a) the output \emph{clus} gives the radius, hydrogen density, temperature, gas pressure, emission measure, turbulent velocity, outflow velocity and relative abundances, and (b) the output \emph{clup} provides the projected radial profile of the number of photons and energy flux, in an energy band which is specified with parameters $E_{\mathrm{min}}$ and $E_{\mathrm{max}}$ (given in units of keV). The \emph{clus} model has in total $80$ different parameters which are summarized in Table\,\ref{Table:all_parameters}.

\subsubsection{Radial density profile}
\label{Sec:density_profile} 

The hydrogen density distribution $n_{\mathrm H}(r)$ is parametrised as the sum of two beta-models \citep{1976A&A....49..137C, 1999ApJ...517..627M}, modified by a possible density jump:
\begin{equation}
	n_{\mathrm H}(r) = \left[ n_1(r) + n_2(r) \right] \times f(r) \;.
	\label{Eq:density_profile}
\end{equation}
Densities $n_i(r)$, where $i = 1,2$, are defined as
\begin{equation}
 n_i(r) = \dfrac{n_{0,i}}{\left[ 1 + (r/r_{c,i})^2 \right]^{1.5 \beta_i}}	\;,
\end{equation}
where $n_{0,1}$ and $n_{0,2}$ are the central densities, $r_{c,1}$ and  $r_{c,2}$ are the core radii, and $\beta_1$ and $\beta_2$ are the slopes of the beta profiles. The optional density jump $f(r)$ is defined as 
\begin{equation}
	\begin{split}
	r &< r_s: f(r) = 1 \;, \\
	r &> r_s: f(r) = \Delta_d (r/r_s)^{\gamma_d} \;,
    \end{split}
\end{equation}
where $r_{s}$ is the discontinuity radius and $\Delta_d$ is the density jump at $r_{s}$. For $\Delta_d > 1$ the density increases outside $r_{s}$ relative to the undisturbed profile, while for $\Delta_d < 1$ it decreases. Further fine-tuning of this thermodynamic discontinuity (which can represent e.g. a shock, or a cold front) can be achieved by changing the power-law parameter $\gamma_d$. If $\gamma_d>0$ ($\gamma_d<0$) the values are increasing (decreasing) at large radii, while $\gamma_d=0$ means a constant jump.

\subsubsection{Radial temperature profile}
\label{Sec:temperature_profile}
The temperature profile $T(r)$ can be written as
\begin{equation}
	T(r) = T_1(r) \times f_1(r) \times f_2(r) \;,
	\label{Eq:temperature_profile}
\end{equation}
where $T_1(r)$ is based on \citet{2006ApJ...640..691V} (Eq.\,6), and is re-written to a different but equivalent form following Eq.\,10 in \citet{2004A&A...413..415K}
\begin{equation}
	\begin{split}
	T_1(r) &= T_c + \left( T_h - T_c \right) \dfrac{(r/r_{tc})^{\mu}}{1+(r/r_{tc})^{\mu}} \;, \\
	f_1(r) &= \frac{(r/r_{to})^{-a}}{\left[ 1 + (r/r_{to})^b \right]^{c/b}}  \;.
	\end{split}
\end{equation}
The central and outer temperatures $T_c$ and $T_h$, respectively, are not the actual temperatures, but the temperatures that would exist without the $f_1(r)$ and $f_2(r)$ terms. The temperature jump $f_2(r)$ is defined as
\begin{equation}
	\begin{split}
		r &< r_s: f_2(r) = 1 \;, \\
		r &> r_s: f_2(r) = \Delta_t (r/r_s)^{\gamma_t} \;,
	\end{split}
\end{equation}
where $\Delta_t$ and $\gamma_t$ are the temperature jump and the power-law parameter, which are defined similarly to the density jump, and the discontinuity radius $r_{s}$ is the same as for the jump in the density profile. We assume that the ion temperature is equal to the electron temperature for each shell.

If one aims to relate the \emph{clus} model temperature parameters to those of \citet{2006ApJ...640..691V}, then 
\begin{equation}
	\begin{split}
		T_c    &= T_{\mathrm{min}} \;, 
		T_h     = T_0  \;, 
		\mu     = a_{\mathrm{cool}}  \;,
		r_{tc} = r_{\mathrm{cool}}  \;,
		r_{to} = r_t  \;,
	\end{split}
\end{equation}
where $T_{\mathrm{min}}$, $T_0$, $a_{\mathrm{cool}}$, $r_{\mathrm{cool}}$, and $r_t$ are parameters from Eq.\,4 and Eq.\,5 in \citet{2006ApJ...640..691V}.

\subsubsection{Radial abundance profile}
\label{Sec:abundance_profile}
The relative metal abundances from hydrogen to zinc (parameters 01 to 30 in the \emph{clus} model) are given in proto-solar units, where the user can specify which set of abundance tables is used (see the SPEX command \emph{abundance}). By default, the reference element which serves as a scaling atom is hydrogen, but it can be also specified by the user. These abundances can be modified by a multiplicative radial scaling law $f_{\mathrm{abu}}(r)$, whose form is a modified version of Eq.\,6 taken from \citet{2017A&A...603A..80M}
\begin{equation}
	f_{\mathrm{abu}}(r) = \dfrac{A}{(1+r/B)^C} \left[ 1-D \exp^{-(r/F)\times (1+r/E)} \right] + G \;,
	\label{Eq:abu_profiles}
\end{equation}
where the default values for parameters ($A=1.34$, $B = 0.021$, $C=0.48$, $D=0.414$, $E=0.163$, and $F=0.0165$) are set to the universal abundance profile values from \citet{2017A&A...603A..80M}. If the abundances are meant to be constant as a function of radius, the user should take care that $f(r)\equiv 1$ for all radii. This can be achieved for instance by setting $A=0$ and $G=1$. The constant term $G$ was not included in \citet{2017A&A...603A..80M}, but may be useful for some applications. The radial scaling works the same way for all chemical elements (i.e. all elements are assumed to have the same radial profile shape); the abundances themselves can be different. The radial scaling is only done for elements with nuclear charge 3 or more (i.e., lithium and higher). Hydrogen and helium are excluded from the scaling. It is therefore highly recommended to keep the reference atom to its default (hydrogen) for this model. 

\subsubsection{Turbulent velocity and radial motions of plasma}
The turbulent velocity $v_{\mathrm{turb}}$ is defined as
\begin{equation}
	v_{\mathrm{turb}}^2 (r) = v_a^2 + \dfrac{v_b |v_b| \times (r/r_v)^2}{1+(r/r_v)^2} \;,
	\label{eq:turbulent_velocity}
\end{equation}
and is equivalent to the root-mean-square velocity $v_{\mathrm{rms}}$, which is related to the micro-turbulent velocity $v_{\mathrm{mic}}$ following a relation
\begin{equation}
	v_{\mathrm{mic}} = \sqrt{2} v_{\mathrm{rms}} = \sqrt{2} v_{\mathrm{turb}} \;.
\end{equation}
The Doppler broadening $\Delta E_D $ of a line with a rest frame energy $E_0$ is then expressed as
\begin{equation}
	\Delta E_D = \dfrac{E_0}{c} \sqrt{ \left( \dfrac{2 k_B T_{\mathrm{ion}}}{A m_p} + v_{\mathrm{mic}}^2 \right)}
\end{equation}
where $c$ is the speed of light, $k_B$ is the Boltzmann constant, $A$ is the atomic weight, $m_p$ is the mass of a proton, and $T_{\mathrm{ion}}$ is the ion temperature. The term $\sqrt{\frac{2 k_B T_{\mathrm{ion}}}{A m_p}}$ represents the thermal broadening. 

At the center, $v_{\mathrm{turb}}$ is given by $v_{\mathrm{turb}} = v_a$, at large distances it is given by $v_{\mathrm{turb}}^2 = v_a^2 + v_b |v_b|$. Positive values of $v_b$ mean increasing turbulence for larger radii, while negative values of $v_b$ mean decreasing turbulence for larger radii. To be more precise, we emphasize that the turbulent velocity as defined in Eq.\,\eqref{eq:turbulent_velocity} refers to the standard deviation of the velocities within each radial bin, including both microscopic and macroscopic gas motions. In addition, the thermal motions of the ion are added in quadrature using the relevant (ion) temperature. Note that for negative values of $v_b$ with $v_b < -v_a$, the turbulent velocity would become imaginary ($v^2<0$). To avoid a crash, SPEX cuts these values off to zero, and continues its calculations.

In addition to turbulence, the model allows for systematic radial motion (inflow or outflow), which is defined as 
\begin{equation}
	v_{\mathrm{rad}} (r) = v_c + \frac{(v_h - v_c) \times (r / r_z)^2}{1+(r / r_z)^2} \;,
\end{equation}
where parameter $v_c$ corresponds to the flow velocity at the core, $v_h$ is the flow velocity at large distances. Positive values of $v_{\mathrm{rad}}$ correspond to the inflow, negative values correspond to the outflow. Radii $r_v$ and $r_z$ are typical radii where the velocity fields ($v_{\mathrm{turb}}$, $v_{\mathrm{rad}}$) change.

\subsubsection{Projection of the radial profiles}
The \emph{clus} model projects the spectra of all shells onto the sky. Instead of the spectrum for the full cluster, users can also specify calculation of the spectrum within a projected annulus, which is given by the inner ($r_{\mathrm{min}}$) and outer ($r_{\mathrm{max}}$) radii given in units of $r_{\mathrm{out}}$. As previously mentioned, at radius $r_{\mathrm{out}}$ the density is set to zero. Thus, for $r_{\mathrm{min}} = 0$ and $r_{\mathrm{max}} = 1$, the full cluster spectrum is obtained. 

If the user wants to calculate the spectrum or profile in a more complex region and not a simple annulus, for instance the projection of a square detector (pixel), the \emph{clus} model provides this option by using the parameter \emph{azim}. For \emph{azim}=0, the full annulus is used, while for \emph{azim}=1, a more complex region can be used. In this latter case, the user must specify the name of an ascii-file through the parameter called \emph{fazi}, which describes the covered azimuthal fraction of the desired region as a function of radius. For a more detailed description of the specific format of this ascii-file, we refer the reader to Sec.\,4.1.6.10 of the SPEX manual\footnote{SPEX manual for \emph{clus} model can be found here: \url{https://var.sron.nl/SPEX-doc/spex-help-pages/models/clus.html}.}.

\subsubsection{Resonance scattering}
\label{Sec:RS}
The resonance scattering (RS) is calculated using Monte Carlo simulations: for the most important spectral lines affected by RS, random photons are drawn per each spectral line. From the 3D emissivity profiles for the relevant lines, $N \times nr$ random photons are drawn, where $nr$ is the number of 3D shells (introduced in Sec.\,\ref{Sec:cluster_model_theory}) and $N$ is an integer which can be set by the user. Each of these individual photons are followed. The calculation for the photon stops when it is either absorbed due to the photo-ionisation continuum (see section 6.2 in \citealp{2008SSRv..134..155K}), or when it leaves the cluster. Alternatively, it can be absorbed and then (a) re-emitted in a new random direction (the resonance scattering), or (b) it decays to a non-ground level, resulting in two or more photons until the atom reaches the ground state again. Such photons are followed until they are destroyed or they escape. At the end of the calculation, statistics is collected on the history of the photon. Note that the \emph{clus} model assumes that all ions are in their ground state when they absorb a photon. Under the assumption of spherical symmetry, the magnitude of RS is the same throughout each spherical shell. The isotropic phase function was assumed for all relevant transitions.

The user can include the calculation of the resonant scattering by setting the parameter \emph{rsca} to unity. There are two available modes for this calculation:
\begin{itemize}[leftmargin=11pt]
	\item mode 1: The number of initial photons is distributed according to the emissivity of individual shells. This mode is more suitable for the overall spectrum of the galaxy cluster, and its accuracy is higher in the cluster cores, from where the bulk of the photons is emitted. 
	\item mode 2: There is the same number of photons in each shell, while the final result is weighted with the shell emissivity. This mode is more accurate (in comparison with mode=1) when analyzing radial line profiles of the resonant lines in the outskirts, where the emissivity is small. 
\end{itemize}

The RS calculations include the most relevant $666$ transitions for H-like to Na-like ions. We use a simple scaling law $N_{\rm hydrogen} = 10^{25} \sqrt{T}$, where $N_{\rm hydrogen}$ is the total hydrogen column density from the core of the cluster to infinity (in units of per square meter), and ${T}$ is the temperature (in keV). This matches approximately the Perseus cluster at a temperature of $4$\,keV. It is based on the simple scaling laws for cluster mass $M$, radius $R$ and density $\rho$:
\begin{equation}
	\begin{split}
        M &\sim R^3\rho \;, ~~~~~\rho = \rm{constant} \;, \\ 
        M &\sim T^{1.5} \;, ~~~~ N_{\rm hydrogen} \sim \rho ~ R \;.	
    \end{split}
\end{equation}
We calculated the optical depths of all ground-state absorption lines in SPEX using the \emph{hot} absorption model for a grid of temperatures $T$ ranging from $0.25$\,keV to $16$\,keV with a step size of a factor of 2 while assuming no turbulence (i.e., the highest possible optical depths) and a column density of $5$ times higher than the scaling relation $N_{\rm hydrogen} = 10^{25} \sqrt{T}$. From this list, we selected all lines with energy larger than $0.1$\,keV, optical depth larger than $0.03$, and optical depth larger than $1$\% of the strongest line of that ion at that temperature. For each resonance line, we only considered the most important decay channels from the excited state (typically stronger than $1$\% of the total decay rate) in order to limit the number of decay routes to the ground to a manageable number. In practice, the number of decay routes was thus limited to a maximum of $7$ (including the main resonance line), and we limited the number of lines to a maximum of $3$ per route (ignoring the lowest energy lines in a few cases, which were always at much lower energy than the X-ray band). Transition energies, oscillator strengths and branching ratios were thus obtained from the current atomic database of SPEX. When this database is updated, the same transitions will be used for the resonance scattering, but branching ratios, oscillator strengths and line energies will automatically be adjusted; these quantities are initialised each time upon the first call of the \emph{clus} model.

If the resonant scattering is included in the SPEX \emph{clus} model calculation, the user has an option to output three diagnostic files with information on the resonant lines. These files have an ascii format and are named: \emph{cluslin1}, \emph{cluslin2} and \emph{cluslin3}. In particular, \emph{cluslin1} includes a summary of all spectral lines where resonance scattering is taken into account, and \emph{cluslin2} and \emph{cluslin3} give the projected properties and 3D radial properties of all the lines summarized in \emph{cluslin1}, respectively. For more details about these ascii files, we refer the reader to the SPEX manual.


\subsection{Sample selection and \emph{clus} model simulations}
\label{Sec:sample_selection}

In Table\,\ref{tab:properties_Vikhlinin_clusters} we provide a sample of objects that were used in this study. The redshift and $r_{500}$ values of A383, A2029, A1795 and A262 were taken from \citet{2006ApJ...640..691V}, for the Perseus galaxy cluster from \citet{2018PASJ...70...10H}, and for the massive elliptical galaxy NGC 4636 from \citet{2015A&A...575A..38P}. We note that this sample is not fully representative of all galaxy clusters, groups nor massive elliptical galaxies, but rather serves as demonstrative example of the capabilities of the \emph{clus} model and its potential applications.

\begin{table*}
	\centering       
	\caption{Redshift $z_{\rm cluster}$ and $r_{500}$ of the clusters used in our study.}
	
	\begin{tabular}{|l|c|c|c|c|c}
		\hline
		Cluster & $z_{\rm cluster}$ & $r_{500}$ [kpc] & $N^{\rm neutral}_{\mathrm H, TOT}$\tablefootmark{a} [cm$^{-2}$] & 1 kpc [$\arcmin$] \\ \hline
		A383 & $0.1883^{[1]}$ & $944^{[1]}$  & $3.88 \times 10^{20}$     & $0.005$    \\ 
		A2029 & $0.0779^{[1]}$ & $1362^{[1]}$ &  $3.70 \times 10^{20}$   & $0.011$  \\ 
		A1795 & $0.0622^{[1]}$ & $1235^{[1]}$ &  $1.24 \times 10^{20}$   & $0.014$  \\ 
		Perseus & $0.01790^{[2]}$ & $1245^{[2]}$ & $20.7 \times 10^{20}$ & $0.046$ \\ 
		A262 & $0.0162^{[1]}$ & $650^{[1]}$ &  $7.15 \times 10^{20}$     & $0.051$  \\ 
		NGC 4636 & $0.0037^{[3]}$ & $350^{[3]}$ & $2.07 \times 10^{20}$  & $0.218$ \\ \hline
	\end{tabular}
	\tablebib{ [1] \citet{2006ApJ...640..691V}; [2] \citet{2018PASJ...70...10H}; [3] \citet{2015A&A...575A..38P}.}
	\tablefoot{\tablefoottext{a}{ $N_{\rm H, TOT}$ is the total hydrogen column density in the direction of the galaxy cluster, which was taken from \citet{2013MNRAS.431..394W}. We used the online tool created by the SWIFT team, which is available here \url{https://www.swift.ac.uk/analysis/nhtot/}.}}
	\label{tab:properties_Vikhlinin_clusters}
\end{table*}

\begin{figure}
	\centering
	\includegraphics[width=0.5\textwidth]{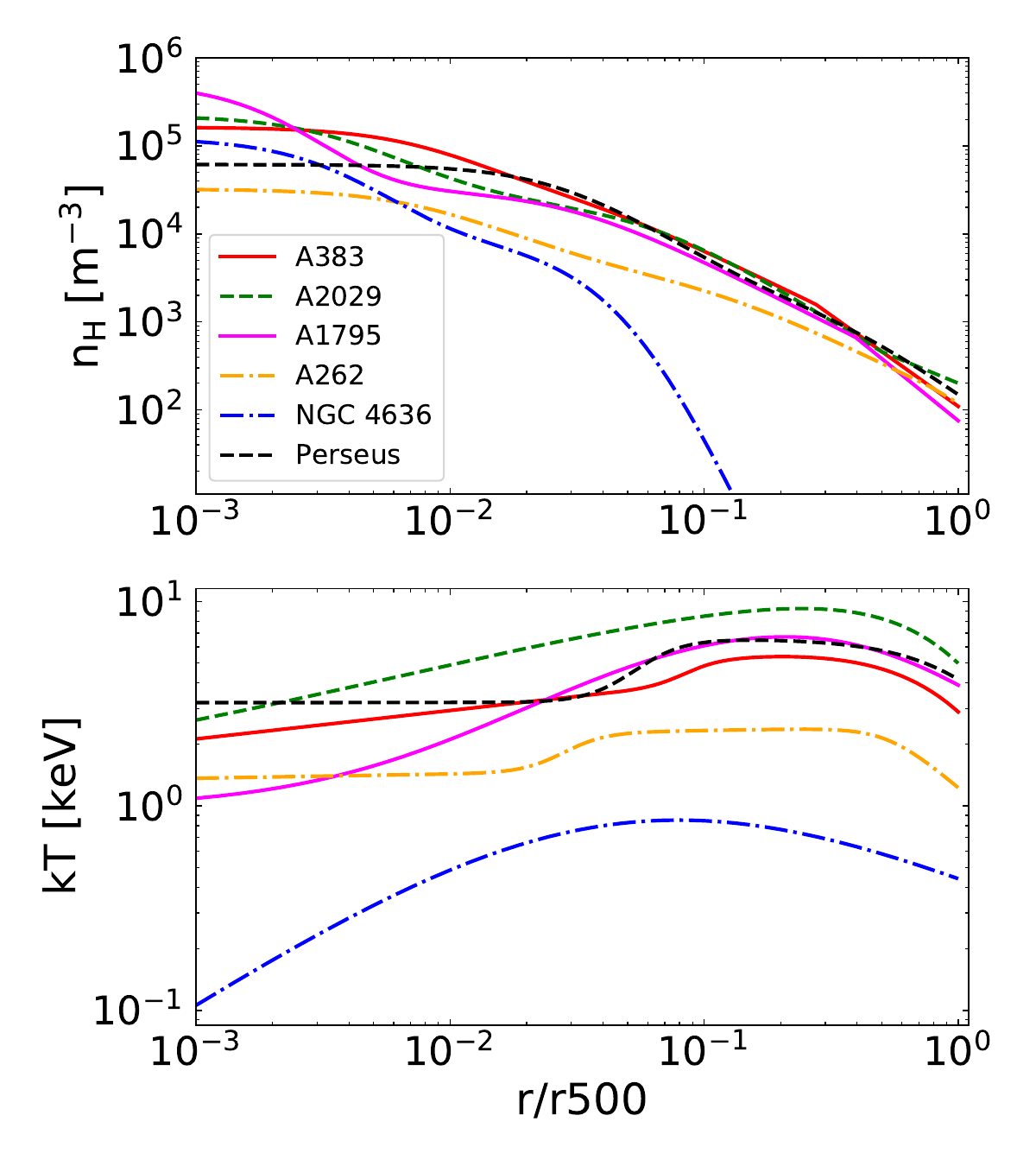}
	\caption{Input radial profiles of hydrogen density $n_{\mathrm{H}}$ (top panel) and temperature $kT$ (bottom panel) for the sample selection from Table\,\ref{tab:properties_Vikhlinin_clusters}. These profiles are obtained by fitting the functional forms of density and temperature profiles as defined in sections \ref{Sec:density_profile} and \ref{Sec:temperature_profile} to the profiles given in \citet{2006ApJ...640..691V} (for galaxy clusters A$383$, A$2029$, A$1795$, and A$262$), to data from \citet{2009ApJ...707.1034B} (for NGC 4636), and data from \citet{2015MNRAS.450.4184Z} (for the Perseus cluster).}
	\label{Fig:density_comparison}
\end{figure}

We simulate spectra of these objects using the \emph{clus} model for two different types of X-ray detectors: (a) ACIS-S CCD (charged coupled device) spectrometer on board of the Chandra X-ray observatory, and (b) Resolve micro-calorimeter on board of the recently launched XRISM observatory. We obtain the input parameters for the density and temperature profiles by fitting the functional form defined in sections \ref{Sec:density_profile} and \ref{Sec:temperature_profile} to the profiles given in \citet{2006ApJ...640..691V} for galaxy clusters A$383$, A$2029$, A$1795$, and A$262$. For NGC $4636$ we use data from \citet{2009ApJ...707.1034B}, and for the Perseus cluster we use data from \citet{2015MNRAS.450.4184Z} corrected for \cite{2009LanB...4B..712L} proto-solar abundances (for more details see section 5.1 in \citealp{2018PASJ...70...10H}). Fig.\,\ref{Fig:density_comparison} shows the comparison of density and temperature profiles for these objects. For the abundance profiles of all objects considered in this study we assume a universal abundance profile taken from \citet{2017A&A...603A..80M}. Such a profile does not have an abundance drop in the innermost few kpc.

In all simulations we leave the turbulent velocity at its default value of $100$\,km/s, and we apply no Poisson noise, which means that the photons in the simulated spectra are not randomised. However, we take the expected uncertainties into account in the determination of the uncertainties on the relevant parameters. This produces correct uncertainties on the parameters, and its advantage lies in the fact that only one simulation is needed, and the result of such simulation can be fully reproducible since the random fluctuations are not taken into account. We also take into account the absorption by our Milky Way (simulated with the \emph{hot} component in SPEX), which is simplified to the absorption by neutral gas only, whose temperature is set to $1  \times 10^{-6}$\,keV. The total hydrogen column densities that we use for the neutral component of the ISM $N^{\rm neutral}_{\rm H, TOT}$ are taken from \citet{2013MNRAS.431..394W} (see Table \ref{tab:properties_Vikhlinin_clusters}). We make two different types of calculations: (a) we neglect the RS effect, and therefore take into account only the projection effects of the gas (including its  multi-temperature and multi-metallicity structure as mentioned in the first paragraph of the following section \ref{Sec:fitting_procedure}), and (b) we add RS to simulations assuming mode 1 (see Sec.\,\ref{Sec:RS}). The parameter $r_{\mathrm{out}}$ is set to $2 \times r_{500}$. In summary, the complete model with which all data in this study is simulated, includes three components: redshift component, hot component and the \emph{clus} model. Unless stated otherwise, all other parameters of the \emph{clus} model are set to their default values in SPEX. 

\subsection{Fitting procedure}
\label{Sec:fitting_procedure} 
The gas in galaxy clusters, groups and massive elliptical galaxies is known to be multi-phase, and unless one fits only the high-energy part of the spectra around the Fe-K complex, single-temperature models can not sufficiently describe the temperature structure in these objects \citep{2000MNRAS.311..176B, 2004A&A...413..415K, 2009A&A...493..409S, 2023arXiv230112791S}. More complex models, which are usually fitted to X-ray spectra, are either a double-temperature model, or the so-called \emph{gdem} model. Some earlier works preferred the \emph{wdem} model instead \citep{2004A&A...413..415K, 2006A&A...449..475W, 2010A&A...523A..81D}. Note that the \emph{clus} model accounts not only for multi-temperature, but also multi-metallicity and multi-velocity structure projected along the LOS. In our results we neglect the multi-velocity structure but we take into account the presence of multiple temperatures and abundances in the 3D profiles of studied clusters. 

A single-temperature model (1T) assumes the gas has just one temperature and that it is in CIE, while a double-temperature model (2T) is a superposition of two CIE models with two different temperatures. The \emph{gdem} model assumes that the emission measure $\mathrm{EM}$($T$)\footnote{$\mathrm{EM}$ is defined as a product of $n_e n_{\mathrm H} d\mathrm{V}$, where $n_e$ is the electron density, $n_{\mathrm H}$ is the hydrogen density, and $d\mathrm{V}$ is the volume filled by the emitting gas.} as a function of temperature $T$ follows a Gaussian distribution, and can be described as
\begin{equation}
	\frac{d ~\mathrm{EM(\textit{x})}}{dx} = \frac{\mathrm{EM}_0}{\sqrt{2\pi} ~ \sigma_{x}} ~ \exp \left( - \frac{ (x - x_0)^2 }{2\sigma^2_x} \right)  \;,
	\label{Eq:gdem_EM_vs_temp}
\end{equation}
where $\mathrm{EM}_0$ is the total, integrated emission measure; $x_0 = \log(k_{\rm B} T_0)$, where $T_0$ is the average temperature of the plasma; and $\sigma_{x}$ is the width of the Gaussian distribution, which in this case is in units of $x = \log(k_{\rm B} T)$. To quantify the discrepancies between these three models (1T, 2T, \emph{gdem}) and the \emph{clus} model, we simulate the spectra following Sec.\,\ref{Sec:sample_selection}, and fit them with 1T, 2T and \emph{gdem} models (including the redshift and \emph{hot} components). For Chandra ACIS-S we use the $2004$ response files, created for the central $100$\,kpc of A$2029$ (Chandra observations ID
4977\footnote{A$2029$ Chandra observations with ID $4977$ can be downloaded from \url{https://doi.org/10.25574/04977}.}). We choose to use the response files from $2004$ observations because the ACIS detectors did not suffer as much from the contamination\footnote{For more details about this contamination of the ACIS detectors or the optical blocking filters, and its effect on quantum efficiency and effective areas see \href{https://cxc.cfa.harvard.edu/ciao/why/acisqecontamN0010.html}{https://cxc.cfa.harvard.edu/ciao/why/acisqecontamN0010.html}.} at low-energies (around and below $1$\,keV). For XRISM Resolve simulated data we use the $2019$\footnote{For more details see \href{https://heasarc.gsfc.nasa.gov/docs/xrism/proposals/index.html}{https://heasarc.gsfc.nasa.gov/docs/xrism/proposals/ index.html}.} response files assuming $5$\,eV spectral resolution. For spectra simulated with Chandra ACIS-S we ignore energies outside the $0.6-7$\,keV energy bins, and for XRISM Resolve we ignore energies outside energy bins between $0.6-12$\,keV while assuming a gate valve open configuration. The data for Chandra ACIS-S CCD are simulated for an exposure time of $500$\,ks, while the XRISM Resolve micro-calorimeter data are simulated for an exposure time of $200$\,ks (except for Fig.\,\ref{Fig:NGC4636_spectrum}, which was simulated with exposure time of $500$\,ks). For all spectra in these energy ranges we use optimal binning following \citet{2016A&A...587A.151K}. We evaluate the goodness of the fit by using C-statistics \citep{1979ApJ...228..939C}, which can be summarised as the maximum likelihood estimation in the limit of Poisson statistics. In SPEX, we use modified C-statistics based on \citet{1984NIMPR.221..437B}, which is described in detail in \citet{2017A&A...605A..51K}. The fit is considered good, if the $\Delta$C-statistics value of the fit is close to the expected C-statistics value. Because we do not include Poisson noise in our spectral simulations, the expected C-statistics for a model exactly matching the one which is used for the simulated data is identical to zero. The C-statistics of each of our fits therefore stem from the mismatches between the exact input model and the best-fit model, and for the rest of this work we will denote this as $\Delta$C-statistics.  
 
For calculations using the 1T model, three parameters are let to vary: normalization, temperature, and the iron abundance. The iron abundance is coupled to abundances from carbon to zinc. The \emph{gdem} model has an additional free parameter $\sigma$, which defines the width of the Gaussian emission measure distribution (see equation \eqref{Eq:gdem_EM_vs_temp}, if $\sigma = 0$, the \emph{gdem} model is identical to 1T model in SPEX). 

For the 2T model we first fit the spectra with the 1T model, and only after finding the best-fit parameters of the 1T model we add the second CIE component. The initial value for the normalization of the second CIE component is assumed to be $10$\% of the best-fit normalization of the first CIE component, and the temperature of the second CIE component is assumed to be $0.5$ times the best-fit temperature of the first CIE component. Only if the normalization of the second component is significant, we free the temperature of the second component. The iron abundance of the first component is coupled to all abundances from carbon to zinc of the first component, and also to the abundances of the second CIE component. In case of the 2T model, the best-fit values of temperature are then weighted by the emission measure (the normalization of CIE models in SPEX is equal to the emission measure). The EM weighted temperature is defined as
\begin{equation}
 T_{\mathrm{EM,weighted}} =  \dfrac{\mathrm{EM}_{\mathrm{CIE1}} \times T_{\mathrm{CIE1}} + \mathrm{EM}_{\mathrm{CIE2}} \times T_{\mathrm{CIE2}}}{ \mathrm{EM}_{\mathrm{CIE1}} + \mathrm{EM}_{\mathrm{CIE2}}}	\;.
\end{equation}
We calculate the error on $T_{\mathrm{EM,weighted}}$ using the error propagation equation. Unless stated otherwise, all errors in this study are reported at the $1\sigma$ level. The best-fit values of the 2T model, for which the second CIE component was not significant (its normalization was zero within 2$\sigma$ level), were removed from our sample.

\section{Results}
\label{Sec:cluster_model_application}

\subsection{Simulations with the Chandra ACIS-S CCD}
\label{Sec:Chandra_simulations}

\begin{figure*}
	\centering
	\includegraphics[width=0.9\textwidth]{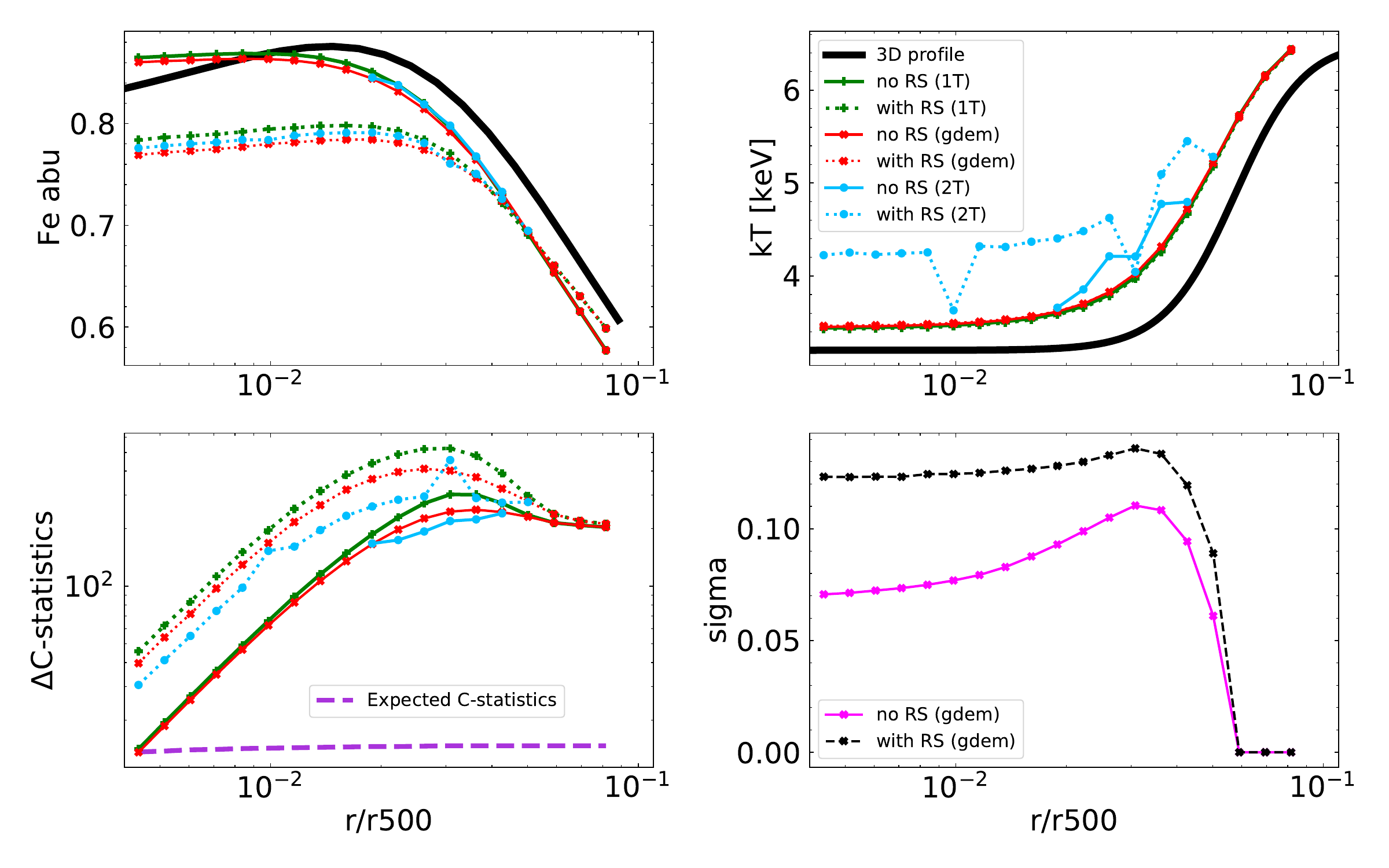}\\
	\caption{Best-fit values of the iron abundance and temperature for spectra of the Perseus cluster simulated with Chandra ACIS-S. Spectra without RS (solid lines) and with RS (dotted lines) are fitted with 1T (green lines), 2T (blue lines), and \emph{gdem} (red lines) models as described in Sec.\,\ref{Sec:fitting_procedure}. The top panels show the best-fit iron abundance values (top left) and best-fit temperature values (top right) together with the 3D profiles (black lines) defined in equations \eqref{Eq:abu_profiles} and \eqref{Eq:temperature_profile}, respectively. The bottom left panel shows the values of $\Delta$C-statistics and the expected value of C-statistics (purple dashed line, see Sec.\,\ref{Sec:fitting_procedure}  for more details). The bottom right panel shows the best-fit value of the $\sigma$ parameter, which is defined in the \emph{gdem} model as the width of the gaussian distribution of the emission measure as a function of temperature. Since the simulated data does not include the Poisson noise, the values of $\Delta$C-statistics which are below the purple curve, indicate an acceptable fit. Spectra for every radial bin were simulated assuming $500$\,ks exposure time. }
	\label{Fig:Chandra_results_Perseus}
\end{figure*}

\begin{figure*}
	\centering
	\includegraphics[width=0.9\textwidth]{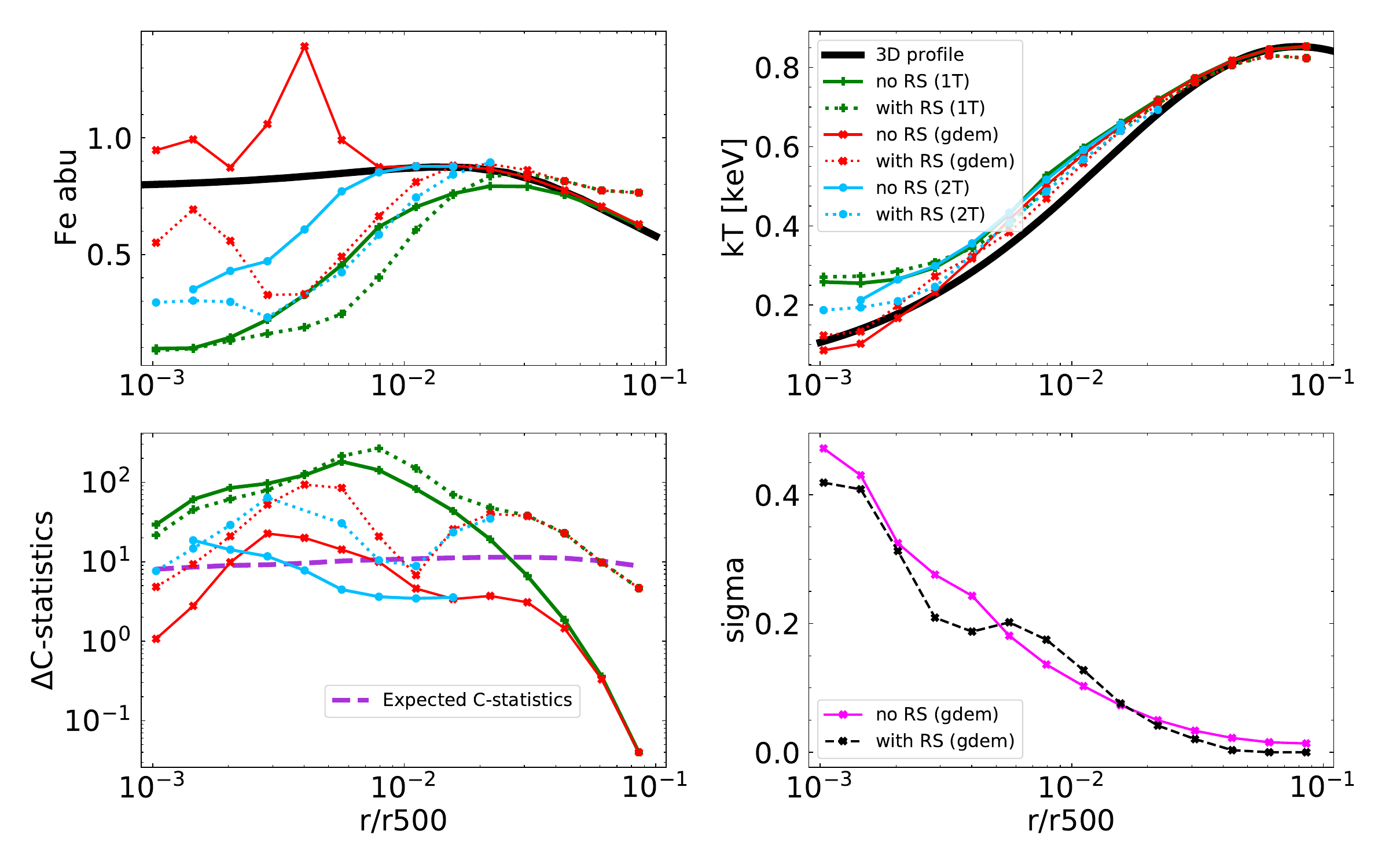}\\
	\caption{Same as Fig.\,\ref{Fig:Chandra_results_Perseus} but for NGC 4636. }
	\label{Fig:Chandra_results_NGC}
\end{figure*}

The angular resolution of Chandra allows us to simulate spectra for the objects in Table\,\ref{tab:properties_Vikhlinin_clusters} for multiple radial bins, and therefore create radial profiles of the iron abundance and temperature. For all objects besides NGC 4636, we define an array of radii between $5-110$\,kpc, which are equally distributed on a logarithmic scale. Due to the low redshift, as well as low mass of NGC 4636, we chose its radial bins between $0.3-35$\,kpc, which are also equally spaced on a logarithmic grid. Our results for the Perseus cluster, NGC 4636, A2029, A1795, A262, and A383 are plotted in Fig.\,\ref{Fig:Chandra_results_Perseus}, Fig.\,\ref{Fig:Chandra_results_NGC}, Fig.\,\ref{Fig:Chandra_results_A2029}, Fig.\,\ref{Fig:Chandra_results_A1795}, Fig.\,\ref{Fig:Chandra_results_A262}, and Fig.\,\ref{Fig:Chandra_results_A383}, respectively. In these figures we plot an upper limit to the expected value of C-statistics (purple dashed line in bottom left panels). The values of $\Delta$C-statistics, which lie below the purple curve, all indicate an acceptable fit.

Fig.\,\ref{Fig:Chandra_results_all} shows the results of fitting 1T, 2T and \emph{gdem} models to spectra simulated with the \emph{clus} model which were extracted for the inner-most and outer-most radial bins (labelled in the caption). We plot the ratio of the best-fit values of the iron abundance and temperature to their input 3D values as defined in equations \eqref{Eq:abu_profiles} and \eqref{Eq:temperature_profile}, respectively. Values of 3D profiles of the iron abundance and temperature are evaluated at the centres of these shells. On the $x$ ($y$) axis we plot the best-fit values to the spectra simulated without (with) RS. Deviations from the vertical line at $x=1$ (along purple arrows in the bottom right panel) indicate the magnitude of projection effects, deviations from the diagonal line $x=y$ (along coral arrows) indicate the magnitude of RS effect, while deviations from the horizontal line at $y=1$ (along teal arrows) indicate a combination of projection and RS effects. Due to the low count rate and large statistical uncertainties, we make the inner-most and outer-most shells of NGC 4636 in Fig.\,\ref{Fig:Chandra_results_all} thicker in comparison with shells chosen for the radial profiles shown in Fig.\,\ref{Fig:Chandra_results_NGC}.

\subsubsection{Temperature profile}
\label{Sec:Chandra_results_temperature}

In the cores of all studied objects, the projected temperature is higher in comparison with the input 3D temperature. For A262, A383, and Perseus this difference is at most by $10-13$\% and for A2029 at most by $20-25$\%. A1795 shows the largest difference out of all studied clusters, where we see an increase in the projected temperature by a factor of $1.5$ (1T), $1.63$ (\emph{gdem}), and $1.8$ (2T) in comparison with the input 3D temperature. The inferred temperatures are significantly higher for $r_{\mathrm{in}}$. This is expected since the 3D temperature profile as a function of radius increases towards smaller radii (before it starts decreasing in the most-inner radius at approximately a few times $10^{-2}$\,$r_{500}$ depending on the mass of the object), and therefore there is hotter gas projected in front of and behind the inner region. However, it is worth noting that, depending on the underlying profiles, some cool-core clusters are affected considerably more than others.

In the outskirts, the projected temperature is higher than its 3D value by at most $5-6$\% for most of the clusters. Perseus shows a slightly higher value of the projected temperature (by $7.5$\%). A262 and NGC 4636 (1T and \emph{gdem}) show a lower projected value of the temperature in comparison with the input 3D value by maximally $4$\%. 

For all galaxy clusters studied in this paper, including or neglecting the RS effect does not make a difference when obtaining the projected temperature profiles in the their cores, but also in their outskirts.

\subsubsection{Iron abundance profile}
\label{Sec:Chandra_results_iron}

As seen from top left panel of Fig.\,\ref{Fig:Chandra_results_all}, the iron abundance of most of the clusters in our sample can be lower or higher by $5-10$\% in their cores due to the projection effects, with an exception of 1T and \emph{gdem} fits to the A1795 spectrum, which shows higher values by approximately $10-20$\% due to the projection effects.  After fitting 1T, 2T and \emph{gdem} models to the simulated spectra of the Perseus cluster, A2029 and A383, we notice similar behaviour of the best-fit iron abundance profiles in their cores. The projected iron abundance profiles show a clear separation between data simulated with and without RS. If constrained, all three models (1T, 2T, and \emph{gdem}) predict lower iron abundance values in the cores of these clusters if the RS is present in their simulated spectra. Contrary to results for the Perseus cluster, A2029 and A383, the projected profiles of the iron abundance in the cores of A1795 and A262 do not show such differences between fitting the data simulated with or without RS. The $\Delta$C-statistics values are worse if RS is taken into account. This result is not surprising since the RS is not included in 1T, 2T, nor \emph{gdem} models.

In the outskirts, the projected iron abundance of the more massive galaxy clusters (A2029, A1795, A383 and the Perseus cluster) in our sample is lower by approximately $10$\% due to the projection effects, while for the less massive object A262 this difference between the projected value and the input iron abundance value rises to $\sim 14-20$\%. The projected values of iron abundance in the outskirts of all clusters in our sample are lower than their 3D input profiles.

\subsubsection{NGC 4636}

\begin{figure*}
	\centering
	\includegraphics[width=\textwidth]{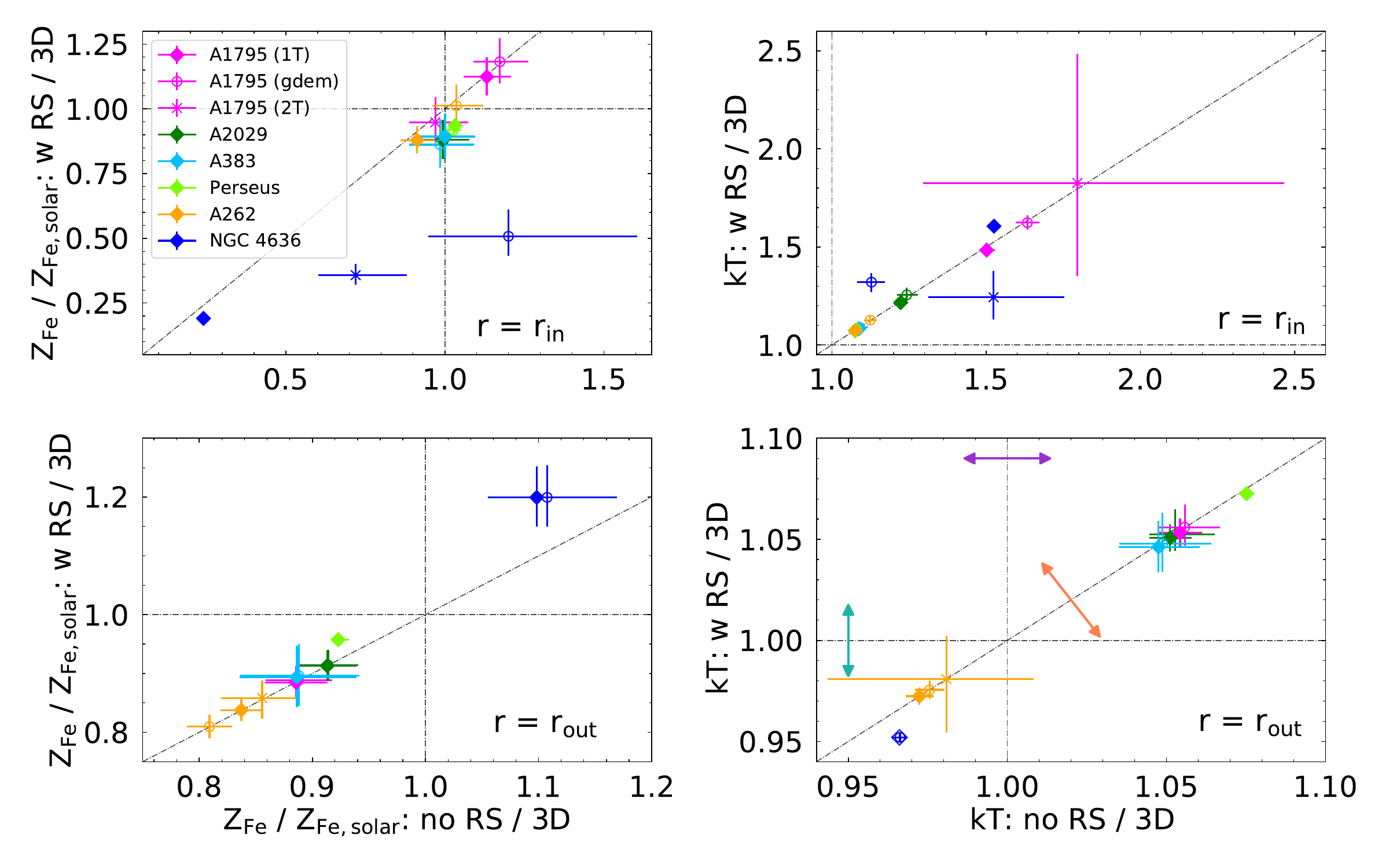}		
	\caption{Best-fit values of 1T (diamond points), 2T (cross) and \emph{gdem} (empty circles) models to the Chandra ACIS-S simulated data with and without the effect of RS for an exposure time of $500$\,ks. The left panels show the best-fit iron abundance values, whereas the right panels show the best-fit values for temperature. The top and bottom panels show the best-fit values for the most inner and the most outer shells, respectively. The shells thickness for most of the objects are approximately between $5-6$\,kpc for $r_{\rm in}$, and $80-110$\,kpc for $r_{\rm out}$, with exception of A383 ($r_{\rm in} \in 5-8.5$\,kpc, $r_{\rm out} \in 66-110$\,kpc) and NGC 4636 ($r_{\rm in} \in 0.3-1.2$\,kpc, $r_{\rm out} \in 12.6-35$\,kpc). All best-fit values are divided by their corresponding input (3D) values. The best-fit values of the 2T model, for which the second CIE component was not significant were removed from this plot (see last paragraph of section \ref{Sec:fitting_procedure}). 1T results of NGC 4636 for kT ($r_{\rm out}$) are plotted with an empty diamond point for visualisation purposes. }
	\label{Fig:Chandra_results_all}
\end{figure*}

The differences between 2D and 3D iron abundance and temperature profiles in the core ($r_{\rm in} \in 0.3-1.2$\,kpc) and outskirts ($r_{\rm out} \in 12.6-35$\,kpc) of NGC4636 are shown in Fig.\,\ref{Fig:Chandra_results_all}. 

The best-fit projected temperature in the core of NGC 4636 is higher by factor $\sim 1.5$ for 1T and 2T models, while the \emph{gdem} model shows smaller difference between 2D and 3D temperatures (an increase by approximately $10$\%). This difference is caused by the projection effects. In the outskirts, the projected temperature is lower by $3.5$\% for 1T and \emph{gdem} model, while the 2T model is statistically not significant. The effect of RS on inferred temperature profile is the largest in the core and in the case of the \emph{gdem} (increase by approximately $18$\%) and 2T models (decrease by approximately $17$\%), and the smallest in case of the 1T model (increase by $\sim 6$\%). In the outskirts, the effect of RS on the projected temperature is within $<2$\%.

The massive elliptical galaxy NGC 4636 shows the biggest differences in the recovered shape of the iron abundance profile. Out of all considered objects in this paper, this is the only case where we can reproduce a drop in the iron abundance profile in the inner-most few kpc, which has been observed with XMM-Newton (see section \ref{Sec:discussion} for more details). Due to projection, the iron abundance in the core drops by $\approx 80$\% (1T) and $\approx 30$\% (2T), respectively. In the case of fitting its simulated spectra with the \emph{gdem} model, we see an increase in the projected value of iron abundance by $\approx 20$\% in comparison with the input iron abundance value. For the 1T model, there is almost no difference between the best-fit iron abundance values with and without RS. On the other hand, if RS is present in the simulated data, the iron abundance inferred from fitting 2T and \emph{gdem} models to simulated spectra shows a decrease by approximately $50$\% and $58$\%, respectively.

\subsubsection{Comparison of the input and fitted emission measure distributions}
\label{Sec:EM_vs_temperature}

\begin{figure*}
	\centering
	\includegraphics[width=0.9\textwidth]{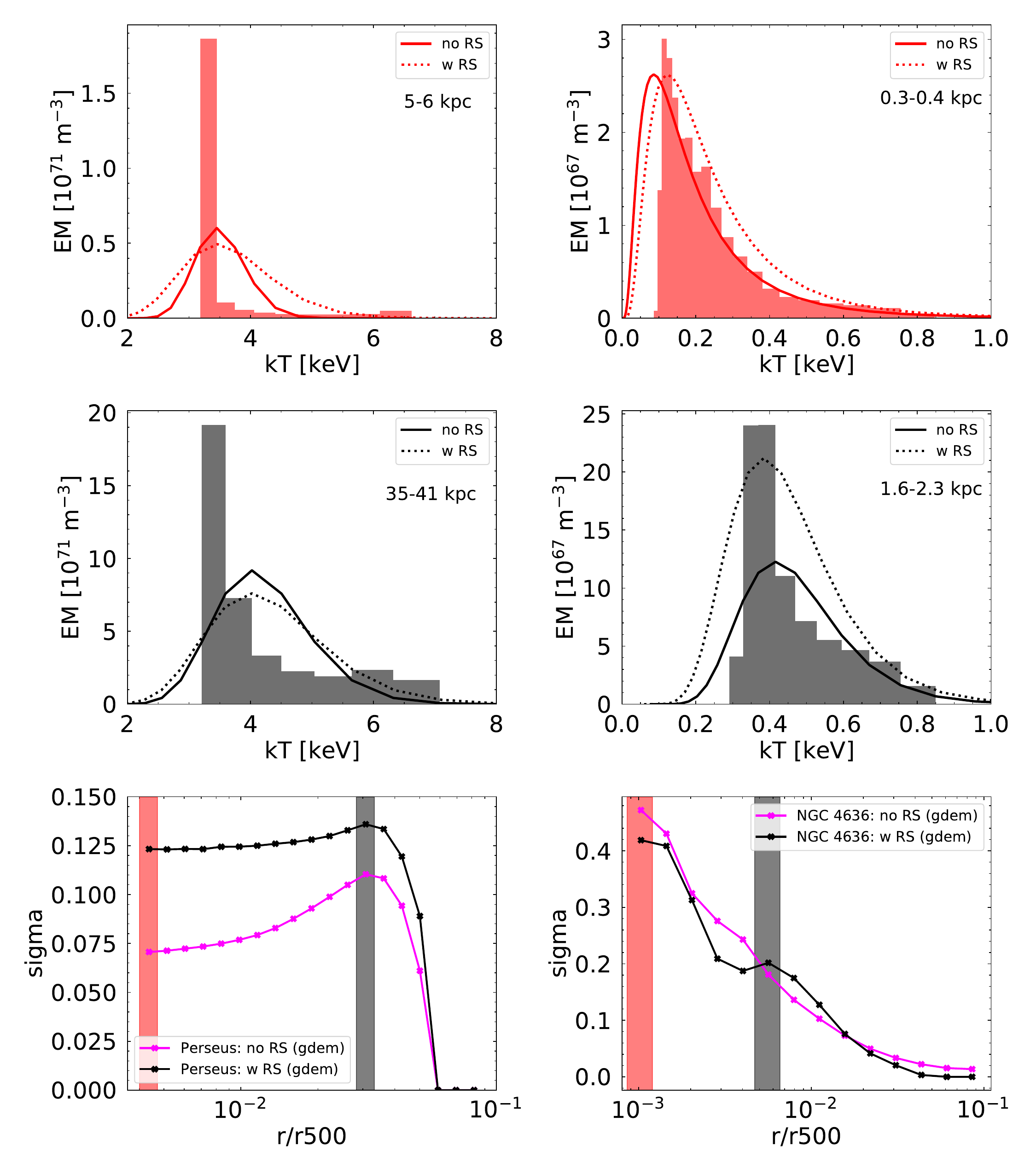}	
	\caption{Emission measure (EM) distribution as a function of temperature for the Perseus cluster (left panels) and NGC 4636 (right panels). Different colours represent different shells which were used to extract the input emission measure distribution as described in Sec.\,\ref{Sec:EM_vs_temperature} (plotted with histograms). The Gaussian functions in each plot were obtained following equation \eqref{Eq:gdem_EM_vs_temp}, and by using the best-fit values obtained and described in more details in Sec.\,\ref{Sec:Chandra_simulations}. The bottom row shows the radial profile of best-fit values of $\sigma$ for the Perseus cluster and NGC 4636, where the red and black bars correspond to radial bins used for deriving the EM distribution in the first two rows of this figure.}
	\label{Fig:EM_vs_temp_plot}
\end{figure*}

In Fig.\,\ref{Fig:EM_vs_temp_plot} we plot the input emission measure as a function of temperature for the Perseus cluster (left panels) and NGC 4636 (right panels), and compare it to the Gaussian distribution retrieved from the best-fit \emph{gdem} model following equation \eqref{Eq:gdem_EM_vs_temp}.  The histograms in Fig.\,\ref{Fig:EM_vs_temp_plot} were created by taking cylindrical cuts through the cluster for different radii (representing 2D/projected shells of the cluster).  For each 3D radius within this cylinder, the 3D density and 3D temperature were calculated using the radial profiles defined in sections \ref{Sec:density_profile} and \ref{Sec:temperature_profile}.  This allowed us to obtain EM in each of these infinitesimally small integration volumes. We then calculated the distribution of EM as a function of temperature in each of these 2D shells and plotted the results with histogram bars. The width of the temperature bins was assumed to be the same as the underlying EM distribution of the best-fit \emph{gdem} model. We inspected two 2D shells while making sure that in one of these 2D shells the sigma parameter reached its maximum (see the bottom right panel in Fig.\,\ref{Fig:Chandra_results_Perseus} and  Fig.\,\ref{Fig:Chandra_results_NGC}).  

Our results show that for both targets, the \emph{gdem} model does not represent accurately the shape of the input EM. The smallest difference between the best-fit \emph{gdem} model and the input EM distribution is for the shells with $\sigma \approx 0.14$ (Perseus) and $\sigma \approx 0.2$ (NGC 4636). Even in this case one might argue that a model which could represent the input EM distribution of these shells more accurately might not be the \emph{gdem} model, but rather a decreasing power-law or a skewed \emph{gdem} model.

In the inner-most projected shell of the Perseus cluster, the input EM is skewed towards a single temperature, whose value coincides with the 3D value in its core ($\approx 3.3$\,keV). Both best-fit \emph{gdem} models (with and without RS) overestimate the amount of gas with temperatures below and above this peak temperature.

In the case of NGC 4636, the best-fit \emph{gdem} model (no RS) peaks at a slightly different temperature in comparison with the input EM distribution, which results in overestimating the amount of gas with temperatures below $0.1$\,keV, and underestimating the gas at higher temperatures. The tail of the EM distribution of this massive elliptical galaxy (above $0.4$\,keV) agrees well with the \emph{gdem} results (no RS). In the case of the best-fit \emph{gdem} values including RS, the \emph{gdem} model peaks at the same temperature as the input EM distribution, however, it overestimates the amount of gas with temperatures lower/higher that this peak. The tail of this \emph{gdem} best-fit distribution agrees well with the input EM distribution for temperatures above $\approx 0.6$\,keV. The input EM distribution could again be described more accurately by a different model, as for example the skewed \emph{gdem} model. 

In the outskirts of both objects, the best-fit \emph{gdem} values of $\sigma$ were either identical with zero, or very close to zero. We would like to stress that the plotted histograms in Fig.\,\ref{Fig:EM_vs_temp_plot} take into account only the effect of projection. When we refer to the gas as being multi-phase, we mean considering its 3D temperature profile as defined in Fig.\,\ref{Fig:density_comparison}, and we neglect other effects that can also change the phase of the gas (e.g. cooling instabilities).

\subsection{Simulations with the XRISM Resolve micro-calorimeter}
\label{Sec:XRISM_simulations}

\begin{figure*}
	\centering
	\includegraphics[width=\textwidth]{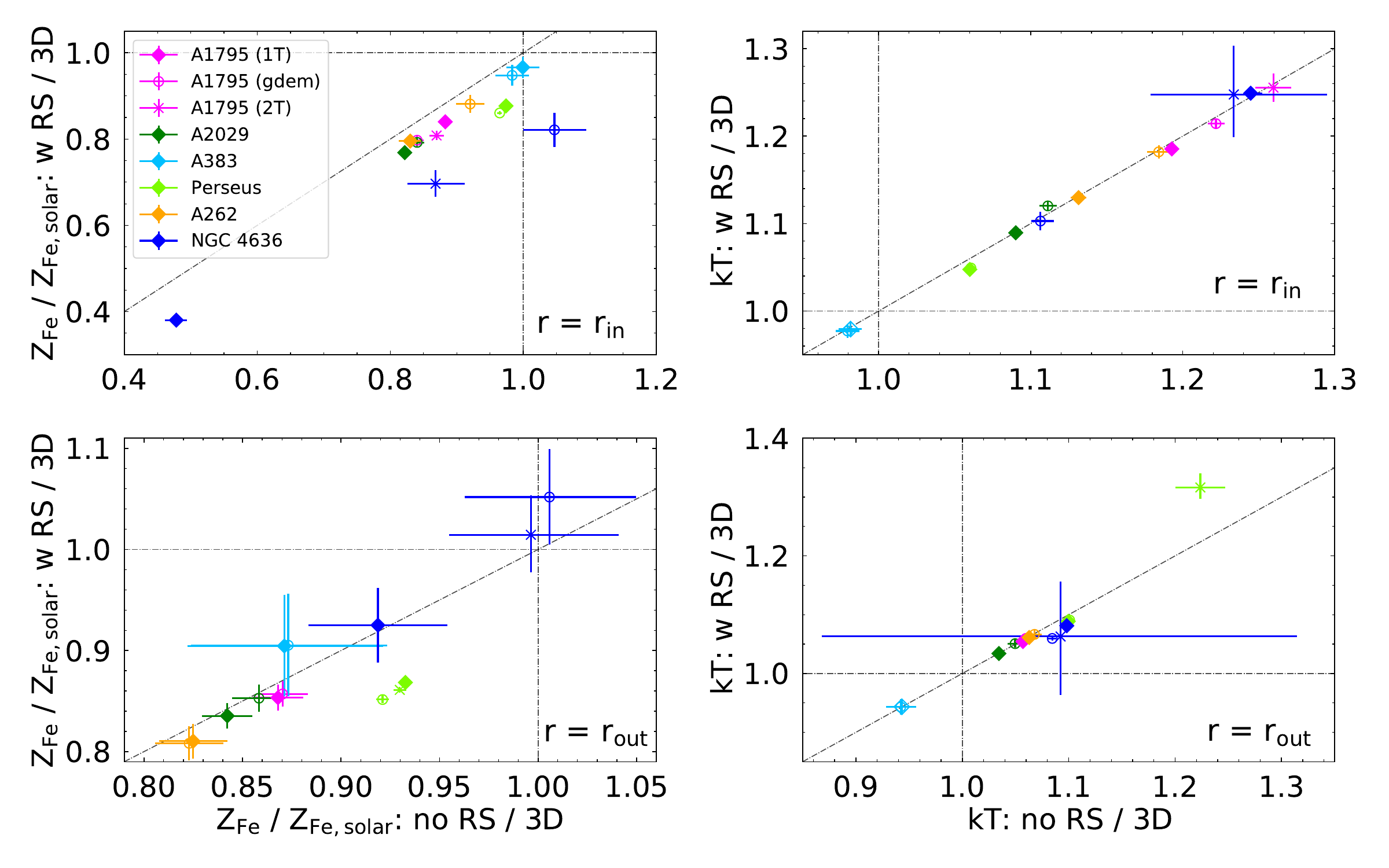}
	\caption{Same as Fig.\,\ref{Fig:Chandra_results_all} but for data simulated with XRISM Resolve micro-calorimeter. The exposure time was assumed to be $200$\,ks. The physical sizes of inner shells are approximately (0-3)\,kpc for NGC 4636, (0-16)\,kpc for Perseus and A262, (0-54)\,kpc for A1795, (0-66)\,kpc for A2029, and (0-142)\,kpc for A383.  The physical sizes of outer shells are approximately (3-7)\,kpc for NGC 4636, (16-33)\,kpc for Perseus and A262, (54-108)\,kpc for A1795, (66-133)\,kpc for A2029, and (142-283)\,kpc for A383. The errors represent the standard deviation $1 \sigma$. Similar to  Fig.\,\ref{Fig:Chandra_results_all}, the best-fit values of the 2T model, for which the second CIE component was not significant, were removed from this plot (see the last paragraph of section \ref{Sec:fitting_procedure}). 1T results of A383 for kT in both radial bins are plotted with an empty diamond point for visualisation purposes.}
	\label{Fig:XRISM_results}
\end{figure*}

The design of the XRISM Resolve micro-calorimeter in combination with its $3'\times 3'$ FoV and the angular resolution of $1.7$ arcminutes limits our simulations to only two radial bins, from which the spectra of studied objects are extracted: 
\begin{itemize}[leftmargin=11pt]
	\item  $r_{\mathrm{in}} = 0-0.75'$, which corresponds to a $1.5'\times 1.5'$ array, and
	\item  $r_{\mathrm{out}} = 0.75'-1.5'$, which corresponds to the rest of the $3'\times 3'$ array. 
\end{itemize}	
In Fig.\,\ref{Fig:XRISM_results} we show the results of fitting 1T, 2T and \emph{gdem} models to simulated spectra extracted for the inner and outer radial bins as defined in the figure caption. We plot the best-fit values of the iron abundance and temperature for two cases: with and without the effect of resonant scattering. The values of 3D profiles of the iron abundance and temperature are evaluated at the centres of these shells following equations \eqref{Eq:abu_profiles} and \eqref{Eq:temperature_profile}, respectively. Our simulations are done assuming the response files for an open gate valve, however, at the moment the gate valve opening attempts have not been successful. In case that the gate valve remains closed for an extended period of time, we plan to redo our simulations in the future with the updated response files.

\subsubsection{Temperature profile}
\label{Sec:XRISM_results_temperature}

In the inner shells, the differences between the 3D temperatures and the best-fit projected values are slightly larger in comparison with the outer shells due to the same reason as we discussed in section\,\ref{Sec:Chandra_results_temperature}. As shown in Fig.\,\ref{Fig:XRISM_results}, the best-fit values for $r_{\mathrm{in}}$ are larger by at most $6$\% (Perseus) to $26$\% (A1795, 2T), with an exception of A383, for which the best-fit temperature is lower by $2$\% in comparison with the input temperature. In the outer shells, the best-fit values are higher at most by $\sim 3$\% (A2029, 1T) to $23$\% (Perseus, 2T) for most of the studied objects. For A383, the 2D temperature is lower by $\sim 6$\%. For all objects besides A383, the best-fit temperatures are always higher than the input 3D temperatures for both inner and outer shells.

From Fig.\,\ref{Fig:XRISM_results} we see that the presence or absence of RS in spectra of almost all objects considered in this study does not affect the projected temperature (within error bars) recovered from fitting simulated spectra with models like 1T, 2T or \emph{gdem} (the data points lie on the diagonal black solid line). This is true for both radial bins ($r_{\mathrm{in}}$ and $r_{\mathrm{out}}$). The only exception is te 2T fit in the outer radial bin of the Perseus cluster, where the projected temperature increased by approximately $7$\% due to the presence of RS in the simulated data.

\subsubsection{Iron abundance profile}

In comparison with Chandra simulations, XRISM simulations yield slightly higher values for the effect of projection, as well as resonant scattering, on the iron abundance in the inner shell of galaxy clusters. The projection effects make the iron abundance in the cores of the galaxy clusters lower by $\sim 1-20$\% when compared to the input 3D values. These effects are the lowest ($\sim 1-5$\%) for the Perseus cluster, A383, and A262 (the \emph{gdem} model). For A2029, A1795, and A262 (1T), the projection effects make the iron abundance in the core of these clusters lower by approximately $8-20$\% in comparison with the input abundance. In the outer shells the projection effects make the inferred iron abundance lower by $13-20$\%, with an exception of the Perseus cluster, for which we see an effect of approximately $6-8$\%.

The effect of RS on the inferred iron abundance is slightly higher in the cores of galaxy clusters in comparison with their outskirts, however, for all clusters studied in this paper we see an effect of maximally $10$\% in both radial bins. Fitting 1T, 2T and \emph{gdem} models to spectra simulated with RS causes a decrease in the projected iron abundance for all clusters from our sample besides the outskirts of A383. The largest effect of RS is in the case of the Perseus cluster ($\sim 11$\% in the core, and $\sim 8$\% in the outer shell). We summarize these results (together with NGC 4636 which is discussed in a separate section) in Table\,\ref{Table:XRISM_abu_results_in_out}. 

\begin{table*}
	\caption{XRISM Resolve best-fit iron abundance\tablefootmark{a} values for 1T, 2T and \emph{gdem} models with (\ding{51}) and without (\ding{55}) RS for the inner (r$_{\mathrm{in}}$) and outer shells (r$_{\mathrm{out}}$) as defined in the caption of Fig.\,\ref{Fig:XRISM_results}.  }
	\begin{tabular}{|l|c|c|c|c|c|c|c|c|c|}
		\hline
		& & \multicolumn{1}{c|}{r$_{\mathrm{in}}$} & \multicolumn{1}{c|}{} & \multicolumn{1}{c|}{} & \multicolumn{1}{c|}{} & \multicolumn{1}{c|}{r$_{\mathrm{out}}$} & \multicolumn{1}{c|}{} & \multicolumn{1}{c|}{} & \multicolumn{1}{c|}{} \\ 
		& RS & \multicolumn{1}{c|}{3D} & \multicolumn{1}{c|}{1T} & \multicolumn{1}{c|}{2T} & \multicolumn{1}{c|}{gdem} & \multicolumn{1}{c|}{3D} & \multicolumn{1}{c|}{1T} & \multicolumn{1}{c|}{2T} & \multicolumn{1}{c|}{gdem} \\ \hline
		A2029 & \ding{55}   & 0.86 & 0.70 $\pm$ 0.01 & -- & 0.72 $\pm$ 0.01 & 0.65  & 0.55 $\pm$ 0.01 & --            & 0.56 $\pm$ 0.01 \\ 
		& \ding{51}   & 0.86  & 0.66 $\pm$ 0.01 & -- & 0.68 $\pm$ 0.01 & 0.65 & 0.54 $\pm$ 0.01 & --          & 0.56 $\pm$ 0.01 \\ \hline
		A1795 & \ding{55}   & 0.86 & 0.76 $\pm$ 0.01 & 0.75 $\pm$ 0.01 & 0.73 $\pm$ 0.01 & 0.68  & 0.59 $\pm$ 0.01 & -- & 0.59 $\pm$ 0.01 \\ 
		& \ding{51}   & 0.86  & 0.73 $\pm$ 0.01 & 0.70 $\pm$ 0.01 & 0.69 $\pm$ 0.01 & 0.68 & 0.58 $\pm$ 0.01 & -- & 0.58 $\pm$ 0.01 \\ \hline
		A383  & \ding{55}   & 0.65  & 0.65 $\pm$ 0.02 & -- & 0.63 $\pm$ 0.02 & 0.41 & 0.36 $\pm$ 0.02 & --         & 0.36 $\pm$ 0.02 \\ 
		& \ding{51}    & 0.65 & 0.62 $\pm$ 0.02 & -- & 0.61 $\pm$ 0.02 & 0.41  & 0.37 $\pm$ 0.02 & --      & 0.37 $\pm$ 0.02 \\ \hline
		Perseus & \ding{55} & 0.85 & 0.83 $\pm$ 0.004 & -- & 0.82 $\pm$ 0.004 & 0.87 & 0.81 $\pm$ 0.003 & 0.80 $\pm$ 0.003           & 0.80 $\pm$ 0.003  \\ 
		& \ding{51} & 0.85 & 0.75 $\pm$ 0.004 & -- & 0.73 $\pm$ 0.004 & 0.87 & 0.75 $\pm$ 0.003 & 0.75 $\pm$  0.003          & 0.74 $\pm$ 0.004 \\ \hline
		A262 & \ding{55}    & 0.87  & 0.72 $\pm$ 0.02 & -- & 0.80 $\pm$ 0.02 & 0.81 & 0.67 $\pm$ 0.01 & --         & 0.67 $\pm$ 0.01 \\ 
		& \ding{51}    & 0.87 & 0.69 $\pm$ 0.01 & -- & 0.77 $\pm$ 0.02 & 0.81  & 0.66 $\pm$ 0.01 & --          & 0.66 $\pm$ 0.01 \\ \hline
		NGC 4636 & \ding{55}& 0.84 & 0.40 $\pm$ 0.01 & 0.73 $\pm$ 0.04 & 0.88 $\pm$ 0.04 & 0.88  & 0.80 $\pm$ 0.03 & 0.87 $\pm$ 0.04 & 0.88 $\pm$ 0.04 \\ 
		& \ding{51}& 0.84  & 0.32 $\pm$ 0.01 & 0.59 $\pm$ 0.03 & 0.69 $\pm$ 0.03 & 0.88 & 0.81 $\pm$ 0.03 & 0.89 $\pm$ 0.03 & 0.92 $\pm$ 0.04 \\ \hline
	\end{tabular}
	\tablefoot{\tablefoottext{a}{The errors are reported at $1 \sigma$ level. The exposure time was assumed to be $200$\,ks. }}
	\label{Table:XRISM_abu_results_in_out}
\end{table*}

\subsubsection{NGC 4636}
Temperature in the core of NGC 4636 inferred from the \emph{gdem} model is affected the least by the projection ($\sim 10$\%), while in the case of 1T and 2T models this value rises to $\sim 24$\%. In the outskirts, all three models yield similar results, and the projection effects cause an increase in temperature by maximally $10$\% in comparison with 3D values. In both radial bins, the RS does not significantly affect the projected temperature obtained from 1T, 2T and \emph{gdem} models.

Similar to our conclusions in Sec.\,\ref{Sec:Chandra_simulations}, the best-fit values of the iron abundance in the core of NGC 4636 are affected the most by fitting 1T, 2T, and \emph{gdem} models to the spectra simulated with the \emph{clus} model. The projection effects make the iron abundance in the core lower by $55$\% and $15$\% for 1T and 2T models, respectively. In the case of the \emph{gdem} model, the projected iron abundance is higher by $5$\% in comparison with the input 3D profile. If the RS is taken into account in the simulated spectrum, the iron abundance decreases by approximately $10$\% (1T), $17$\% (2T), and $22$\% (\emph{gdem}). In the outskirts, the 1T model shows the largest sensitivity to the projection effects, while the \emph{gdem} model shows the largest sensitivity if RS is taken into account. Both of these effects are below $10$\%.

\begin{figure}
	\centering
	\includegraphics[width=\columnwidth]{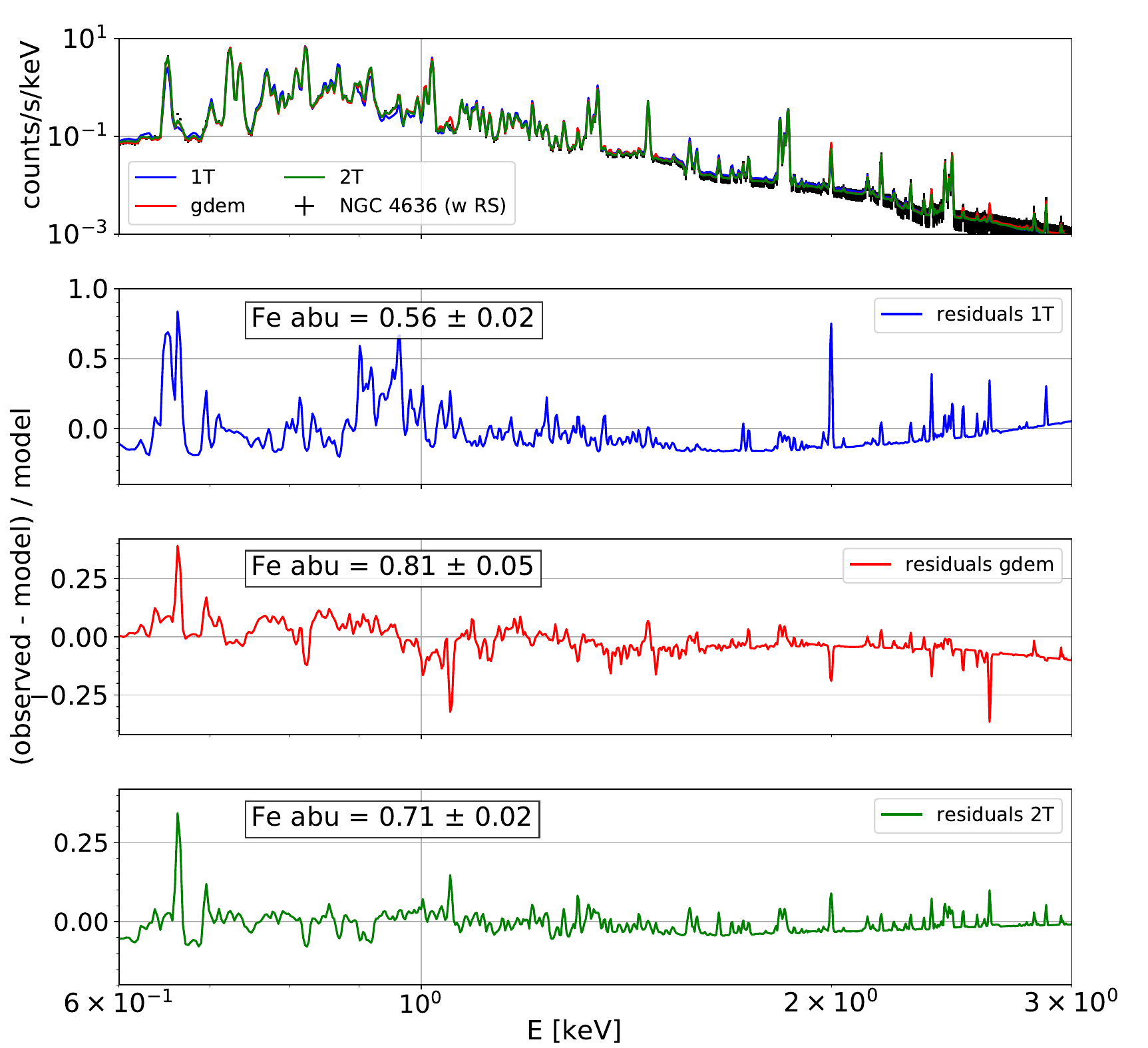}
	\caption{XRISM Resolve spectra in the energy range $0.6-3$\,keV simulated with the \emph{clus} model (including RS) for the $0.3-7$\,kpc aperture region of NGC 4636 (black points), while assuming the gate valve is open. The best-fit 1T, 2T and \emph{gdem} models are plotted with blue, green and red solid lines together with the residuals which are shown in bottom three panels. The best-fit values of the iron abundance are given in the plots together with 1$\sigma$ errors. The exposure time was assumed to be $500$\,ks. }
	\label{Fig:NGC4636_spectrum}
\end{figure}

In Fig.\,\ref{Fig:NGC4636_spectrum} we show NGC 4636 simulated spectra (with RS) for a shell from $0.3$-$7$\,kpc (full FOV) for $500$\,ks XRISM/Resolve observation. We compare data simulated with the \emph{clus} model (black points) to the best-fit 1T (blue solid line), 2T (green solid line), and \emph{gdem} (red solid line) models. We plot the residuals between the simulated data and the 1T, 2T and \emph{gdem} models in the bottom three panels. These residuals show that 1T, 2T and \emph{gdem} models can not fully describe the richness of the NGC 4636 spectra, especially if the resolution of the observed spectrum is similar to XRISM Resolve. Additionally, the discrepancy between 2D and 3D profiles for XRISM simulated data is slightly lower in comparison with simulations for Chandra, which is mostly due to the difference in their spatial resolution.

\section{Discussion}
\label{Sec:discussion}

\subsection{Multi-temperature and multi-metallicity structure of the ICM}
Our results show that even if the temperature and gas density distribution is smooth, and at every shell the gas is isothermal, while no thermal instabilities are present in the ICM, the projection effects influence the best-fit temperatures obtained from fitting the complex spectra of galaxy clusters and NGC 4636 with models such as 1T, 2T, and \emph{gdem}. In Fig.\,\ref{Fig:Chandra_results_all} we showed that in cores of some cool-core clusters the projected temperature is within $8-12$\% from the input temperature (see e.g. core of A262), while in other cool-core clusters, as for example the core of A1795, the projection effects caused a difference between the input temperature and the projected temperature which is as large as a factor of $1.5-1.8$. In case of the core of NGC 4636 (1T and 2T models) this factor is $1.5$, while the temperature inferred from the \emph{gdem} model is $10$\% higher in comparison with the input value. For all objects studied in this paper, the 2D temperature is higher in comparison with the 3D input temperature.

The temperature in the inner radius inferred from XRISM simulations is less affected by the projection effects (for most of the clusters) in comparison with Chandra simulations (see Fig.\,\ref{Fig:XRISM_results}). On the other hand, the temperature inferred from XRISM simulations in $r_{\mathrm{out}}$ is affected slightly more by projection in comparison with Chandra simulations. 

It is well known that the 1T model can not describe the multi-temperature structure of galaxy clusters accurately, and leads to an ``iron bias"  \citep{2000MNRAS.311..176B} for the low-temperature objects, and to an ``inverse iron bias" for the high-temperature objects \citep{2008ApJ...674..728R, 2009A&A...493..409S, 2010A&A...522A..34G}. \citet{2009A&A...493..409S} concluded that for the cool-core cluster Hydra A, one needs a broad emission measure distribution such as the \emph{gdem} model to fit simultaneously the lines of the Fe-K and Fe-L complexes, which indicate a multi-temperature origin of the gas. E.g. \citet{2004A&A...413..415K} concluded that the 2T model can in most cases sufficiently describe the spectra of cool-core clusters. However, as our results show, neither 2T nor \emph{gdem} model can in many cases accurately describe the underlying emission measure distribution as shown in Fig.\,\ref{Fig:EM_vs_temp_plot}, even when the fit is statistically acceptable (see all values of $\Delta$C-statistics which lie below the purple dashed curve in bottom left panels of Fig.\,\ref{Fig:Chandra_results_NGC}, Fig.\,\ref{Fig:Chandra_results_A2029}, Fig.\,\ref{Fig:Chandra_results_A1795}, Fig.\,\ref{Fig:Chandra_results_A262}, and Fig.\,\ref{Fig:Chandra_results_A383}). This shows the difficulty of reconstructing the input temperature profile from the projected values, even for a smooth temperature profile assuming a set of isothermal shells and neglecting any cooling instabilities. Such findings might impact the cross-calibration studies between different instruments, which in return might affect the inferred masses of galaxy clusters, and therefore also the cosmological parameters (see e.g. \citealp{2022MNRAS.517.5594W, 2024A&A...688A.107M}).

\subsection{Measured Fe abundance drop in the cores}

As we already mentioned in the introduction of this paper, some galaxy clusters, galaxy groups, and massive elliptical galaxies show a sharp decrease in the radial profiles of the metal content in their inner-most few kiloparsecs. A unified theory explaining the reason behind this abundance drop has not yet been established. In this paper, we investigated whether projection effects and the resonant scattering process can potentially explain the measured abundance profiles in some of these objects.

Out of the objects studied in this paper, only the cores of the Perseus cluster and the massive elliptical galaxy NGC 4636 have a confirmed abundance drop in their projected profiles, e.g. in the XMM-Newton observations (\citealp{2009MNRAS.398...23W} for NGC 4636, and \citealp{2017A&A...603A..80M} for NGC 4636 and the Perseus cluster). We compare our results to the work of \citet{2017A&A...603A..80M} since the authors used the same set of proto-solar abundances by \cite{2009LanB...4B..712L}. The iron abundance was measured by fitting \emph{gdem} to the XMM-Newton data, resulting in a decrease from $0.9$\,$Z_{\rm Fe}/Z_{\rm Fe, \astrosun}$ at $\sim 0.02\,r_{500}$ to $0.5$\,$Z_{\rm Fe}/Z_{\rm Fe, \astrosun}$ at $\sim 0.003\,r_{500}$ for NGC 4636, while for the Perseus cluster the iron abundance drops from $0.8$\,$Z_{\rm Fe}/Z_{\rm Fe, \astrosun}$ at $\sim 0.03\,r_{500}$ to $0.6$\,$Z_{\rm Fe}/Z_{\rm Fe, \astrosun}$ at $\sim 0.004\,r_{500}$. The authors find no abundance drop in A2029 and in A1795, while A262 shows a sharp increase from $0.6$\,$Z_{\rm Fe}/Z_{\rm Fe, \astrosun}$ at $\sim 0.1\,r_{500}$ to $1.4$\,$Z_{\rm Fe}/Z_{\rm Fe, \astrosun}$ at $\sim 0.007\,r_{500}$. Similar conclusions of no significant abundance drop in the inner-most few kpc were reached for A2029 with Chandra ACIS-S observations (Figure 4 of \citealp{2002ApJ...573L..13L}), for A1795 with XMM-Newton observations (Figure 1 of \citealp{2001A&A...365L..87T}) and with Chandra ACIS observations (Figure 7 of \citealp{2002MNRAS.331..635E}), and for A262 with Suzaku observations (Figure 3 of \citealp{2009PASJ...61S.365S}). Besides a global abundance value reported for A383 in \citet{2010ApJ...713..491M}, which reaches $ 0.52 \pm 0.07$ of the solar value for \citet{1998SSRv...85..161G} abundances, we could not find the radial profile of the iron abundance in currently existing literature. 

A few previous works such as \citet{2007MNRAS.380.1554R} or \citet{2008MNRAS.390.1207R} concluded that projection effects do not affect the measured iron abundance profiles significantly. Depending on the mass of the simulated objects, as well as the projected distance from the core, we find different conclusions from our simulations. 

\subsubsection{Perseus}
For the core of the Perseus cluster \citet{2004MNRAS.347...29C} and \citet{2004ApJ...600..670G} concluded that there is no evidence for resonant scattering based on XMM-Newton data. However, measurements with the Hitomi SXS micro-calorimeter showed that the resonant scattering suppresses the flux measured in the \ion{Fe}{XXV} He$\alpha$ line by a factor of $\sim 1.3$ in the inner $\sim 30$\,kpc \citep{2018PASJ...70...10H}. From our simulations with Chandra ACIS-S as well as XRISM Resolve we can conclude that
\begin{itemize}[leftmargin=11pt]
	\item the projection effects can not explain the abundance drop in the core of the Perseus cluster, and all used models (1T, 2T, and \emph{gdem}) cause an increase (decrease) in the iron abundance maximally by $5$\% based on simulations with Chandra (XRISM). 
	\item the effect of resonant scattering can only partly explain the iron abundance drop in the core of Perseus, and according to our results, the measured iron abundance can maximally decrease by $10-15$\%. 
\end{itemize}
Similar conclusions were reported in \citet{2006MNRAS.370...63S}, who showed that resonant scattering can not remove the central drop in the Chandra CCD spectra of galaxy clusters (namely Centaurus and A2199), and that this effect can change the measured metallicities at most by $10$\%. These results are in agreement with our studies. Additionally, the Perseus cluster is not a perfectly symmetric system, and faint X-ray cavities, as well as weak shocks,  ripples, and radio lobes around its central galaxy NGC 1275 are present in its ICM (see e.g. \citealp{1996AJ....112...91M, 2011MNRAS.418.2154F, 2018PASJ...70....9H}). Such effects are, however, neglected in our simulations. We defer the simulations of non-spherically symmetric systems as well as additional clusters known to host abundance dips to future work.

\subsubsection{NGC 4636}
The massive elliptical galaxy NGC 4636 is known for showing resonant scattering in its core, which changes the flux ratio between the resonant line at $15.01\AA$ and forbidden lines at $17.05\AA$ and $17.10\AA$ of \ion{Fe}{XVII} \citep{2002ApJ...579..600X, 2003ASPC..301...23K, 2009PASJ...61.1185H, 2009MNRAS.398...23W, 2016A&A...592A.145A}. As we mentioned earlier in this paper, it also has a steep abundance drop measured in its core, which makes it a perfect candidate to test the theory whether accounting for the RS effect can remove this central abundance drop.

NGC 4636 was one of the five massive elliptical galaxies studied in \citet{2009MNRAS.398...23W}. The authors fitted the observed spectra with a 1T model, and concluded that more complicated models did not improve the fit, and that differential emission measure models always converged to the single-temperature approximation. This is not in agreement with our findings in Fig.\,\ref{Fig:EM_vs_temp_plot}, which shows that in the core, the differential emission measure should follow a decreasing power-law or perhaps a skewed \emph{gdem} model. 

Depending on the model (1T, 2T, or \emph{gdem}) which we used to fit the simulated data of NGC 4636, we confirm that the central abundance drop in this massive elliptical galaxy can be potentially explained by RS, and in the case of the 1T model also solely by projection effects. The models like 1T, 2T, and \emph{gdem} neglect the RS effect, which influences the measurement of fluxes in \ion{Fe}{XVII} lines. Our results indicate larger differences between 3D and projected values ($50-58$\% depending on the model) in comparison with those reported in \citet{2009MNRAS.398...23W}, where the authors concluded that RS can lead to underestimating Fe and O abundances by $10-20$\% in its core. 

As previously mentioned, in \citet{2017A&A...603A..80M} the iron abundance measured using the \emph{gdem} model drops from $0.9$\,$Z_{\rm Fe}/Z_{\rm Fe, \astrosun}$ at $\sim$ $ 0.02\,r_{500}$ to $0.5$\,$Z_{\rm Fe}/Z_{\rm Fe, \astrosun}$ at $\sim 0.003\,r_{500}$. Using the same model as the authors (\emph{gdem}) to fit the spectra we simulated with the \emph{clus} model, while taking into account RS, we report a drop approximately from $0.86$\,$Z_{\rm Fe}/Z_{\rm Fe, \astrosun}$ at $\sim 0.02\,r_{500}$ to $0.33$\,$Z_{\rm Fe}/Z_{\rm Fe, \astrosun}$ at $\sim 0.003\,r_{500}$. These findings are in agreement with observations by \citet{2017A&A...603A..80M} (within statistical uncertainties).  

Similar to our discussion of the results for the Perseus cluster, we would like to point out that NGC 4636 is known for its non-spherical nature. Its X-ray images show a presence of X-ray bubbles and cavities, as well as spiral arms, which are believed to originate from shocks caused by the central AGN, and its interaction with the surrounding ICM (see e.g. \citealp{2002ApJ...567L.115J, 2009ApJ...707.1034B}). The \emph{clus} model assumes a spherical symmetry when simulating X-ray spectra. This assumption breaks down for NGC 4636, which might affect the results reported in this paper. However, we do not expect that the central abundance drop for this object would completely disappear.


\section{Conclusions}
\label{Sec:conclusions}

This paper introduces the \emph{clus} model, which was recently implemented in the software package SPEX. The \emph{clus} model can be used for any X-ray emitting plasma which is in CIE, and which can be approximated with spherical symmetry. The advantage of this model lies in the forward modelling of spectra and radial profiles of selected sources, assuming their 3D temperature, density, velocity and metal abundance profiles. The X-ray emitting gas is divided into a set of spherically symmetric shells, where the emission in each shell is described with a single CIE model, and projected onto the sky. This model also includes the resonant scattering (RS) effect, which is implemented through the Monte Carlo simulations. 

We used the \emph{clus} model to simulate spectra of galaxy clusters Abell 383, Abell 2029, Abell 1795 and Abell 262, the Perseus cluster, and the massive elliptical galaxy NGC 4636. We modelled spectra of these objects with and without resonant scattering while assuming CCD-like (Chandra ACIS-S) and micro-calorimeter (XRISM Resolve) spectral resolutions. We created projected radial profiles of their metal abundance and temperature by fitting their simulated spectra with models like the single-temperature (1T), double-temperature (2T), and the Gaussian-shaped differential emission measure (\emph{gdem}) models.

As shown in this paper, the impact of projection and RS effects vary depending on the mass/temperature of the object, as well as the projected distance from its core. Our main conclusions can be summarized as follows:
\begin{itemize}[leftmargin=11pt]	
	\item As shown by the differential emission measure for the Perseus cluster and NGC 4636 (see Fig.\,\ref{Fig:EM_vs_temp_plot}), the 1T, 2T, or \emph{gdem} models are not always accurately describing the underlying EM distribution. The EM distribution changes depending on the projected distance from the cluster core as well as the thickness of the shell. Depending on these properties, different models such as e.g. skewed \emph{gdem} model, or decreasing power-law might be more suitable to describe the EM profile. Hence fitting data with a model, which does not represent the underlying distribution, can lead to fitting biases.
	
	\item The projection effects cause an increase of temperature inferred from fitting 1T, 2T, or \emph{gdem} models to the simulated spectra in the cores of studied objects, with the exception of XRISM simulations for A383. For clusters with less prominent core in their density profile, this increase is $10-30$\% ($6-18$\%) for Chandra (XRISM), while for the objects with a more prominent core in the density profile, such as A1795 and NGC 4636, this increase can be as high as a factor of $1.5-1.8$ if spectra were simulated with Chandra. For XRISM simulations, these factors are slightly lower $\sim (1.2-1.26)$ due to lower spatial resolution. In the outskirts of galaxy clusters, the differences are below $8$\% for Chandra, however, XRISM simulations show that the projected temperature can be higher by $3-23$\% in comparison with its input value.

	\item Using models like the \emph{clus} model for fitting CCD and micro-calorimeter spectra is more crucial for obtaining the abundance profiles of low-mass and low-temperature objects (see Fig.\,\ref{Fig:Chandra_results_all} and Fig.\,\ref{Fig:XRISM_results}). In the outskirts of A262, the projected iron abundance is $14-20$\% lower than the input profile irrespective of the model used for fitting its simulated spectra. In the core of the massive elliptical galaxy NGC 4636, the projected iron abundance is lower by almost $80$\% ($55$\%) for 1T model, or higher by $20$\% ($5$\%) for the \emph{gdem} model if spectra were simulated with Chandra (XRISM). Regardless of the instrument, the \emph{gdem} model is affected the least by the projection effects in the core of NGC 4636. However, we report non-negligible differences between the input and projected profiles also for hotter and more massive objects. For some cool-core clusters, the differences might be negligible (e.g. iron abundance in the core of A383 simulated with Chandra and XRISM), while in other cool-core clusters as for example A1795, the projected iron abundance in its core is approximately $20$\% higher (lower) in comparison with the input profile, if data was simulated with Chandra (XRISM). 
	
	\item The resonant scattering can reduce the observed central abundance drop in galaxy clusters such as the Perseus cluster by maximally $10-15$\%, but it can not fully explain the abundance drop in the inner-most few kpc. Similar conclusions were reported in \citet{2006MNRAS.370...63S} for Centaurus and A2199 galaxy clusters. 
	
	\item In case of the massive elliptical galaxy NGC 4636, the resonant scattering can explain the abundance drop measured in its core. Depending on the model used for fitting its spectra, not accounting for the resonant scattering leads to an underestimation of the iron abundance in the core of this massive elliptical galaxy by $\sim 50$\% (2T) to $58$\% (\emph{gdem}), if data was simulated with Chandra. In the case of the 1T model, this abundance drop can also be solely explained by projection effects. The XRISM simulations show that RS can make the iron abundance in the core of NGC 4636 lower by maximally $10-22$\%. Our results indicate larger differences between 3D and projected values ($50-58$\% depending on the model) in comparison with those reported in \citet{2009MNRAS.398...23W}, where the authors concluded that RS can lead to underestimating Fe and O abundances by $10-20$\% in its core.
	 
\end{itemize}

\begin{acknowledgements} 
The authors acknowledge the financial support from NOVA, the Netherlands Research School for Astronomy. A.S. acknowledges the Kavli IPMU for the continued hospitality. SRON Netherlands Institute for Space Research is supported financially by NWO. L.S. acknowledges the financial support of the GA\v{C}R EXPRO grant No. 21-13491X.
\end{acknowledgements}

\section*{Data availability}
The dataset generated and analysed during this study will be available in the ZENODO repository. 

\bibliography{bibliography.bib} 


\begin{appendix}

\section{Additional material}
In Table\,\ref{Table:all_parameters} we provide a complete set of all \emph{clus} parameters. The acronyms of the parameters are their names as used in the SPEX interface and all variables are described in Sec.\ref{Sec:cluster_model_theory}. 

\begin{table}[hbt!]
	\caption{The complete set of  $80$ parameters of the \emph{clus} model. }
			\centering
			\resizebox{\textwidth}{!}{{%
			\begin{tabular}{|c|c|c||c|c|c|}
				\hline
				\textbf{Acronym} & \textbf{Variable} & \textbf{Units} & \textbf{Acronym} & \textbf{Variable} & \textbf{Units} \\ \hline
				r500 & $r_{500}$ & 1e+22 m & dfe & D & -- \\ 
				rout & $r_{\mathrm{out}}$ & $r_{500}$ & efe & E & $r_{500}$ \\ 
				nr & Number of 3D shells & --  & ffe & F & $r_{500}$ \\ 
				npro & Number of projected annuli & -- & gfe & G & -- \\
				hd1 & $n_{0,1}$ & m$^{-3}$ & av & $v_a$ & km/s \\ 
				rc1 & $r_{c,1}$ & $r_{500}$ & bv & $v_b$ & km/s \\
				bet1 & $\beta_{1}$ & -- & rv & $r_v$ & $r_{500}$ \\ 
				hd2 & $n_{0,2}$ & m$^{-3}$ & zc & $v_c$ & km/s \\ 
				rc2 & $r_{c,2}$ & $r_{500}$ & zh & $v_h$ & km/s \\ 
				bet2 & $\beta_{2}$ & -- & rz & $r_z$ & $r_{500}$ \\ 
				rsh & $r_{s}$ & $r_{500}$ & emin & $E_{\mathrm{min}}$ & keV \\ 
				dfac & $\Delta_{d}$ & -- & emax & $E_{\mathrm{max}}$ & keV \\ 
				dgad & $\gamma_{d}$ & -- & rmin & $r_{\mathrm{min}}$ & $r_{\mathrm{out}}$ \\ 
				tc & $T_c$ & keV & rmax & $r_{\mathrm{max}}$ & $r_{\mathrm{out}}$ \\ 
				th & $T_h$ & keV & azim & -- & -- \\ 
				rtc & $r_{tc}$ & $r_{500}$ & fazi & -- & -- \\ 
				mu & $\mu$ & -- & rsca & Resonant scattering & -- \\ 
				rto & $r_{to}$ & $r_{500}$ & nit0 & N & -- \\ 
				at & a & -- & rmod & Mode of resonant scattering & -- \\ 
				bt & b & -- & out1 & Output 1 & -- \\ 
				ct & c & -- & out2 & Output 2 & -- \\ 
				tfac & $\Delta_{t}$ & -- & out3 & Output 3 & -- \\ 
				dgat & $\gamma_{t}$ & -- & ref & Reference element & -- \\ 
				afe & A & -- & 01 … 30 & Abundances H to Zn in proto-solar units & -- \\ 
				bfe & B & $r_{500}$ & file & Filename for non-thermal electron distribution (optional) & -- \\ 
				cfe & C & $r_{500}$ & & & \\ \hline
			\end{tabular}
			}}
	\label{Table:all_parameters}
\end{table}
\FloatBarrier

\section{Chandra simulations for other clusters}
\begin{figure*}
	\centering
	\includegraphics[width=0.95\textwidth]{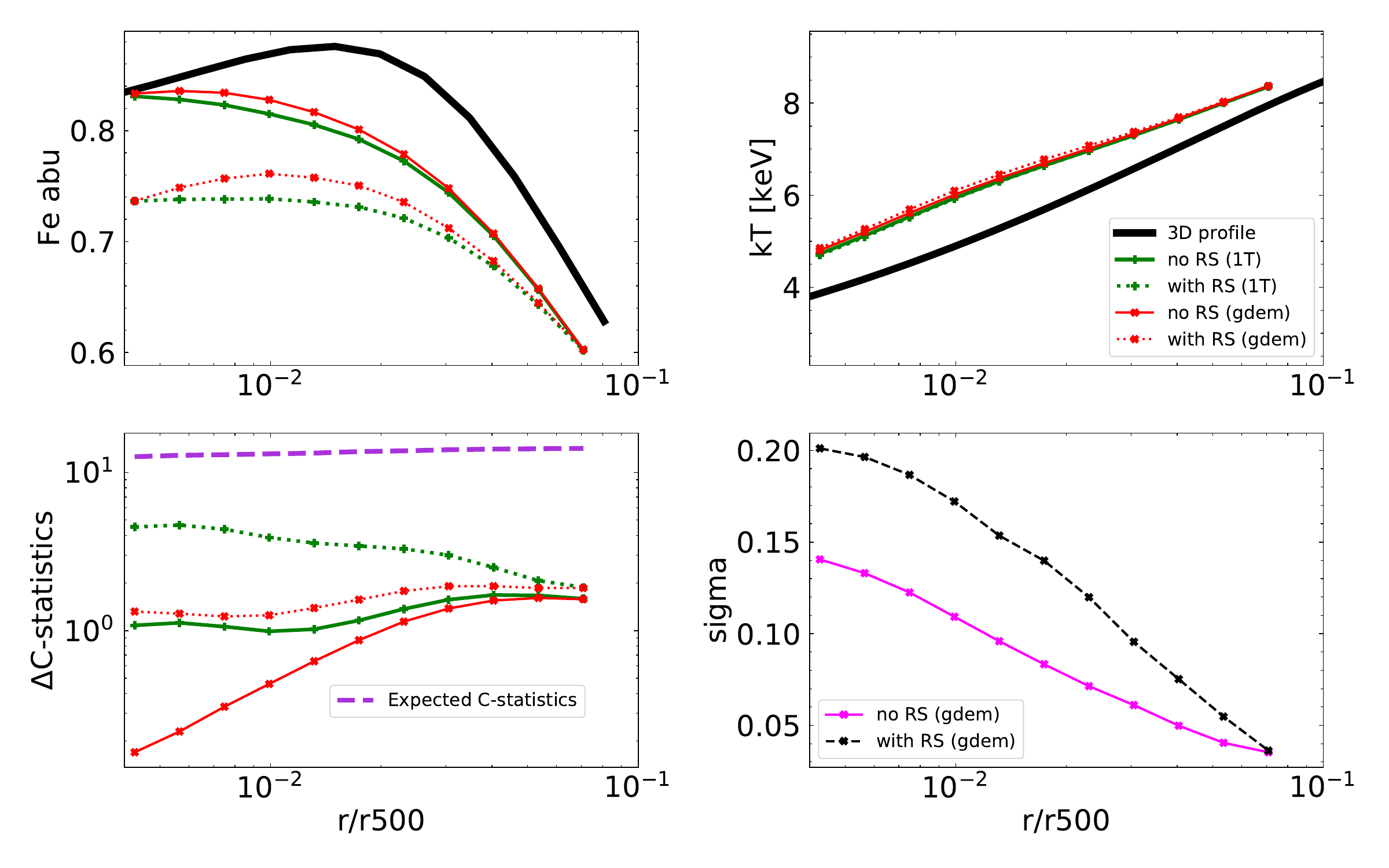}\\
	\caption{Same as Fig.\,\ref{Fig:Chandra_results_Perseus} but for A2029.}
	\label{Fig:Chandra_results_A2029}
\end{figure*}

\begin{figure*}
	\centering
	\includegraphics[width=0.95\textwidth]{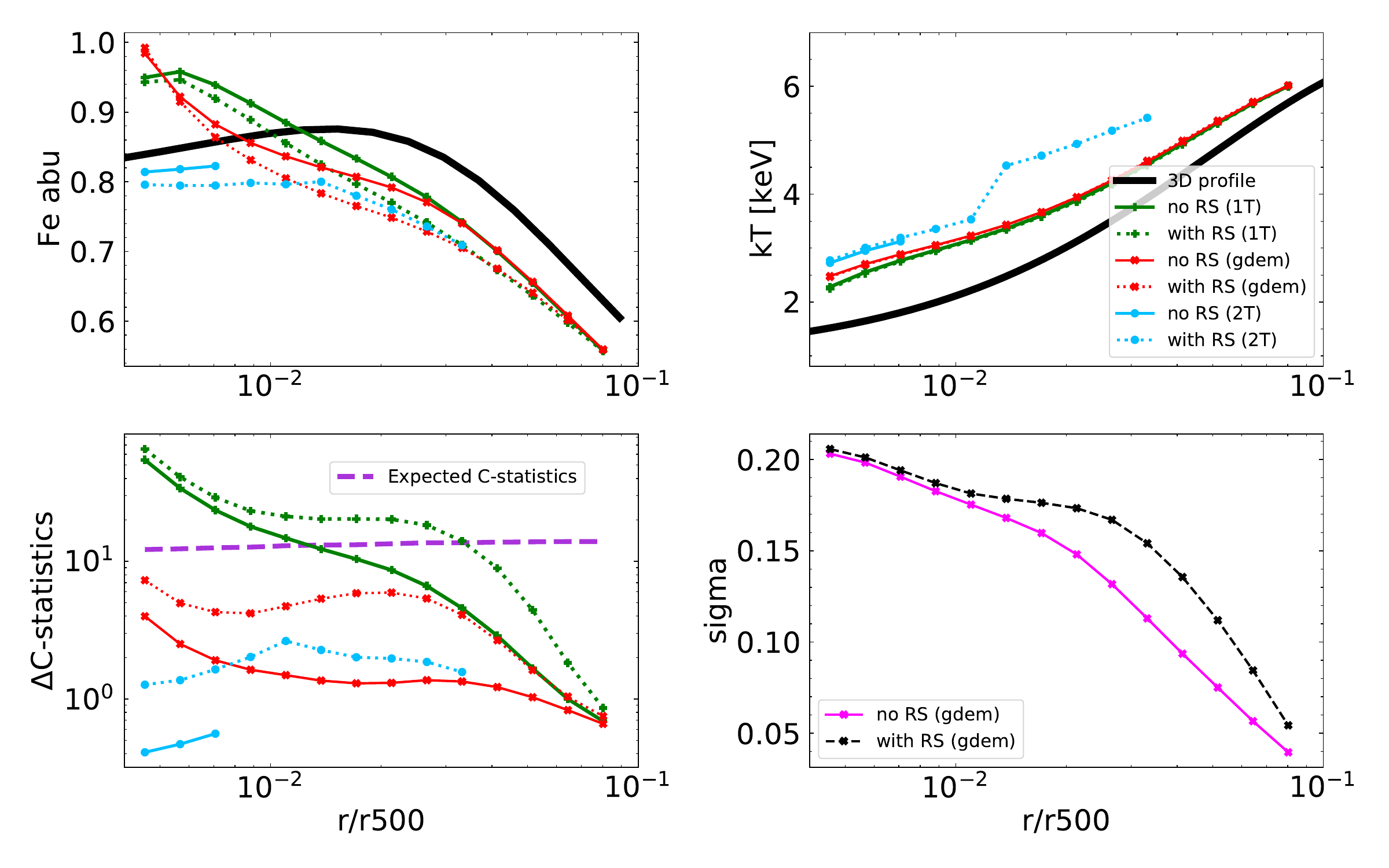}\\
	\caption{Same as Fig.\,\ref{Fig:Chandra_results_Perseus} but for A1795.}
	\label{Fig:Chandra_results_A1795}
\end{figure*}

\begin{figure*}
	\centering
	\includegraphics[width=0.95\textwidth]{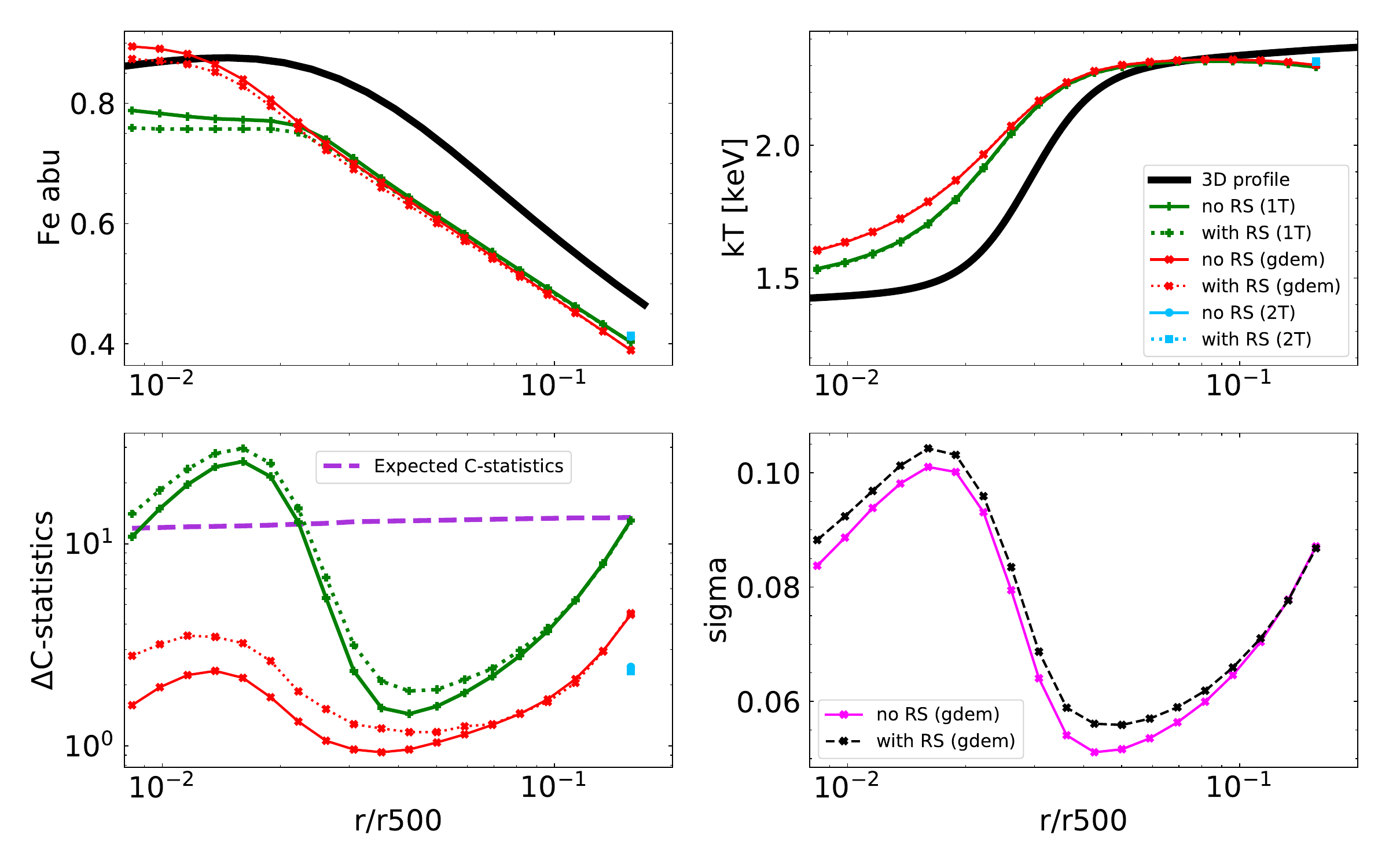}\\
	\caption{Same as Fig.\,\ref{Fig:Chandra_results_Perseus} but for A262.}
	\label{Fig:Chandra_results_A262}
\end{figure*}

\begin{figure*}
	\centering
	\includegraphics[width=0.95\textwidth]{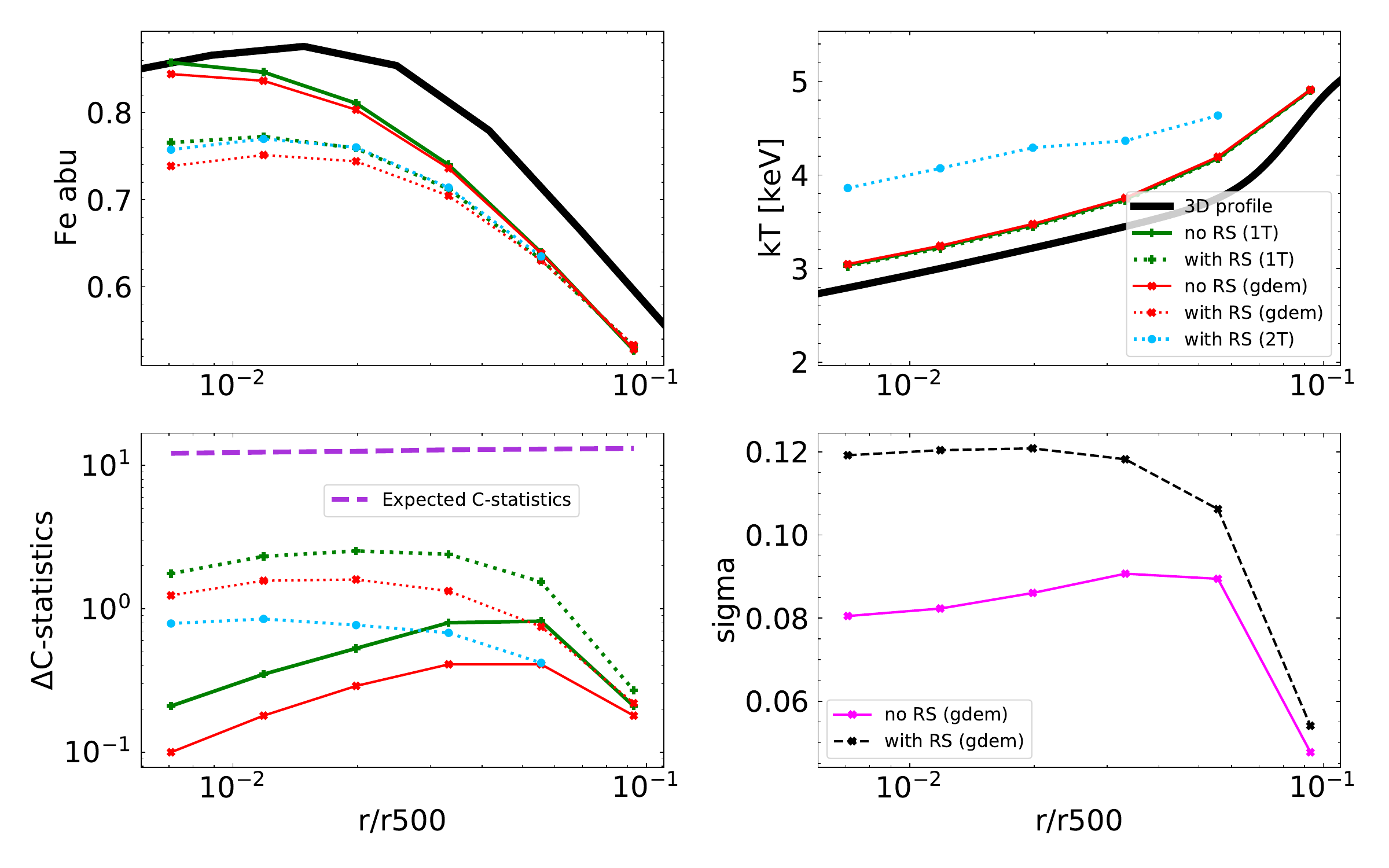}\\
	\caption{Same as Fig.\,\ref{Fig:Chandra_results_Perseus} but for A383.}
	\label{Fig:Chandra_results_A383}
\end{figure*}

\end{appendix}

\label{lastpage}
\end{document}